\documentclass[aip,prx,reprint]{revtex4-1}

\usepackage{graphicx}
\usepackage{dcolumn}
\usepackage{bm}
\usepackage{amsmath}
\usepackage{amssymb}
\usepackage{gensymb}
\usepackage{array}
\usepackage{enumitem}
\usepackage{dcolumn}
\usepackage{xcolor}
\usepackage{soul}
\usepackage{listings}
\usepackage[capitalise]{cleveref}
\newcommand*{\pdfscale}{0.8} 
\definecolor{codegreen}{rgb}{0,0.6,0}
\definecolor{codegray}{rgb}{0.5,0.5,0.5}
\definecolor{codepurple}{rgb}{0.58,0,0.82}
\definecolor{backcolour}{rgb}{0.95,0.95,0.92}
\lstdefinestyle{mystyle}{
    backgroundcolor=\color{backcolour},   
    commentstyle=\color{codegreen},
    keywordstyle=\color{magenta},
    numberstyle=\tiny\color{codegray},
    stringstyle=\color{codepurple},
    basicstyle=\ttfamily\footnotesize,
    breakatwhitespace=false,         
    breaklines=true,                 
    captionpos=b,                    
    keepspaces=true,                 
    numbers=left,                    
    numbersep=5pt,                  
    showspaces=false,                
    showstringspaces=false,
    showtabs=false,                  
    tabsize=2,
    escapeinside = @
}
\draft 

\begin{document}


\title{Singlet fission spin dynamics from molecular structure: a modular computational pipeline} 



\author{Dominic M. Jones}
\affiliation{ARC Centre of Excellence in Exciton Science, School of Physics, University of New South Wales, Sydney, New South Wales 2052, Australia}

\author{Thomas MacDonald}
\affiliation{ARC Centre of Excellence in Exciton Science, School of Physics, University of New South Wales, Sydney, New South Wales 2052, Australia}

\author{Timothy W. Schmidt}
\affiliation{ARC Centre of Excellence in Exciton Science, School of Chemistry, University of New South Wales, Sydney, New South Wales 2052, Australia}

\author{Dane R. McCamey}
\email{dane.mccamey@unsw.edu.au}
\affiliation{ARC Centre of Excellence in Exciton Science, School of Physics, University of New South Wales, Sydney, New South Wales 2052, Australia}


\date{\today}

\begin{abstract}
Singlet fission, which has applications in areas ranging form solar energy to quantum information, relies critically on transitions within a multi-spin manifold. These transitions are driven by fluctuations in the spin-spin exchange interaction, which have been linked to changes in nuclear geometry or exciton migration. Whilst simple calculations have supported this mechanism, to date little effort has been made to model realistic fluctuations which are informed by the actual structure and properties of physical materials. In this paper, we develop a modular computational pipeline for calculating singlet fission spin dynamics by way of electronic structural calculations, molecular dynamics, and numerical models of spin dynamics.  The outputs of this pipeline aid in the interpretation of measured spin dynamics and allow us to place constraints on geometric fluctuations which are consistent with these observations.

\end{abstract}


\maketitle 

\section{\label{sec:intro}Introduction}
Singlet fission (SF) is the rapid, spin conserving process in which one singlet excitation splits into two triplet excitations of lower energy. This requires that the energy of one singlet state in a given material is approximately matched to the energy of two triplet states.\cite{Smith2010SingletFission} SF has the potential to increase the thermodynamic efficiency of solar energy harvesting devices via exciton downconversion~\cite{Hanna2006SolarAbsorbers, Tayebjee2012}, while the ability to optically access spin-polarised excitons on fast timescales\cite{Tayebjee2017QuintetFission} has been investigated as a route to dynamic nuclear polarization (photo-DNP)\cite{Kawashima2022} and in quantum information.\cite{Jacobberger2022UsingCrystals} Exploiting SF-derived spin through these applications requires that we understand the origins of these unique spin properties and the relationship of spin functionality to molecular structure. In this work, we demonstrate a computational pipeline for modelling SF spin dynamics commencing from a molecular structure.

\begin{figure}[ht!]
    \includegraphics[scale=\pdfscale]{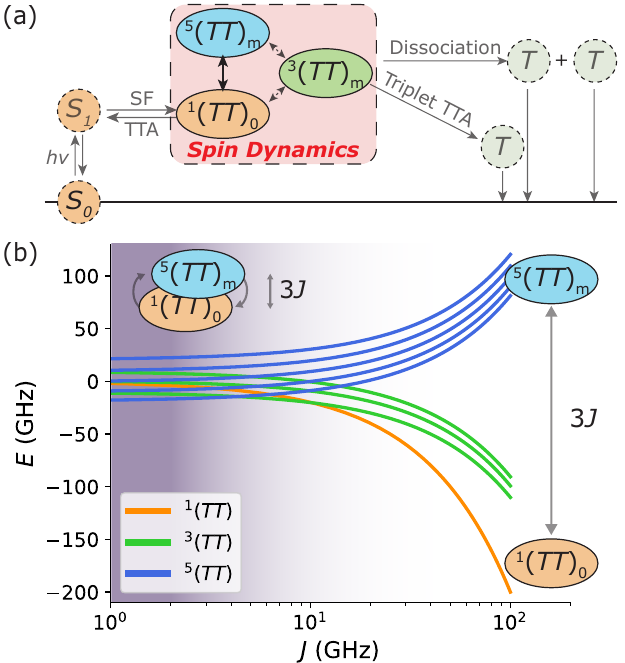}
    \caption{(a) In singlet fission (SF) a singlet exciton decays to an $S=0$ triplet-pair, $^1(TT)_0$, which may dephase to access different spin states $^{2S+1}(TT)_m~(S=0,1,2)$. 
    (b) Energies of triplet pair spin states as a function of exchange coupling $J$.
    When $J$ is small, spin states are poorly defined and mix freely (shaded area) but at large $J$ spin mixing does not occur.}
    \label{fig:Simple SF}
\end{figure}

\Cref{fig:Simple SF} (a) shows a basic model of singlet fission where photoexcited singlet $S_1$ decays into a triplet-pair multiexciton $^1(TT)_0$\cite{Merrifield1969FissionCrystals,Merrifield1971MagneticInteractions}. 
The $^1(TT)_0$ multiexciton can then dissociate to free triplets $(T+T)$ and/or dephase to $^{2S+1}(TT)_m$ spin states of triplet or quintet multiplicity. Another $(TT)$ loss pathway is triplet-channel annihilation, $^3(TT)_m\rightarrow{}^{3}T_m$, where triplet multiexcitons annihilates non-radiatively to give triplet excitons.\cite{He2022} Transient electron spin resonance studies (tr-ESR) have found  $^5(TT)_m$ quintets ($S=2$) as precursors to free triplet formation,\cite{Tayebjee2017QuintetFission,Weiss2017StronglySemiconductor} and it has been suggested that high yields of quintets may reduce energy losses in SF photovoltaics by inhibiting multiexciton recombination\cite{Chen2019Quintet-tripletTemperature}. 

The spin Hamiltonian for the coupled triplet pair $^{2S+1}(TT)_m$ can be written as the sum of Zeeman interaction $\hat{H}_\textrm{Z}$, zero-field splitting term $\hat{H}_\textrm{zfs}$, and intertriplet exchange interaction $\hat{H}_\textrm{ee}$~\cite{Merrifield1969FissionCrystals,Merrifield1971MagneticInteractions}
\begin{subequations}
\begin{eqnarray}
    &\hat{H}& = \hat{H}_\textrm{Z} + \hat{H}_\textrm{zfs} +\hat{H}_\textrm{ee}, \label{eq:Sham}
    \\
    &\hat{H}_\textrm{ee}& = J\hat{S}_1\cdot\hat{S}_2,
\end{eqnarray}
\end{subequations}
where $J$ is the isotropic exchange integral describing single electron exchange between the two triplet pairs---effectively, a measure of through-bond or through-space contact between the two triplets\cite{Kollmar1982TheorySplittings}.

Spin dephasing of the $^{2S+1}(TT_m)$ multiexciton is highly sensitive to the magnitude of the $J$ exchange term. Spin mixing to form quintets is only efficient when the exchange term $J$ is small relative to zero field splitting parameter $D$ (\ref{fig:Simple SF} (b), left),\cite{Collins2019FluctuatingFission,Weiss2017StronglySemiconductor} but the well-resolved $^5(TT)$ quintets observed in tr-ESR are consistent with a $J$ that is large at the time of measurement  (\ref{fig:Simple SF} (b), right). These seemingly contradictory requirements for $J$ to be both small (for $^5(TT)_m$ formation) and large (consistent with experiment) have led to suggestions that $J$ must fluctuate over time\cite{Weiss2017StronglySemiconductor,Tayebjee2017QuintetFission,Nagashima2018,Collins2019FluctuatingFission,Kobori2020GeometriesFissions,Collins2022QuintetFission}, which is supported by recent theoretical works showing the ability of simple $J(t)$ models to drive high quintet yields over experimentally relevant timescales.\cite{Collins2019FluctuatingFission,Kobori2020GeometriesFissions,Collins2022QuintetFission} 

Building on recent findings\cite{Collins2019FluctuatingFission,Kobori2020GeometriesFissions,Collins2022QuintetFission} that SF quintets can form with abstracted models of $J(t)$, we now extend this approach to calculate SF spin dynamics for molecular materials undergoing molecular dynamics calculated from physical models.

\section{\label{sec:Method}Methodology}

\begin{figure}
    \includegraphics[scale=\pdfscale]{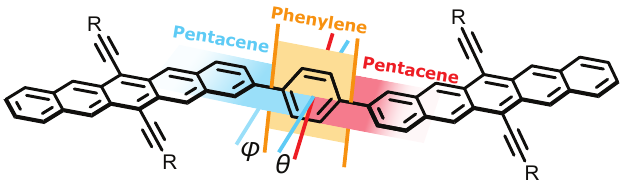}
    \caption{Modelled pentacene dimer \textbf{BP1}. Thermally accessible ring rotations occur about carbon-carbon single bonds and are fully parametrised by a two-dimensional geometry coordinate $\bm{X}(\theta,\phi)$. For real \textbf{BP1} $\mathrm{R = Si(^iPr)_3}$; here $\mathrm{R = H}$ to reduce computational complexity.}
    \label{fig:DoF_BP1}
\end{figure}

We introduce a modular process for simulating the singlet fission spin dynamics of molecular materials. This model provides an end-to-end pathway from electronic structure calculations to fundamental spin physics, bridging the disciplinary gap between chemistry and spin physics by using molecule-specific properties as input parameters for numerical simulations of spin dynamics. We apply this model to the spin dynamics of a well known pentacene dimer, \textbf{BP1}\cite{Trinh2017}(\cref{fig:DoF_BP1}). \textbf{BP1} is known to undergo singlet fission\cite{Trinh2017} 
and has two principle degrees of internal rotational freedom due to the single (1,4)-phenylene bridge. We anticipate that either of these rotations will modulate $J$ \emph{via} changes in exciton delocalisation.\cite{Collins2019FluctuatingFission, Ringstrom2022} 

\begin{figure}
    \includegraphics[scale=\pdfscale]{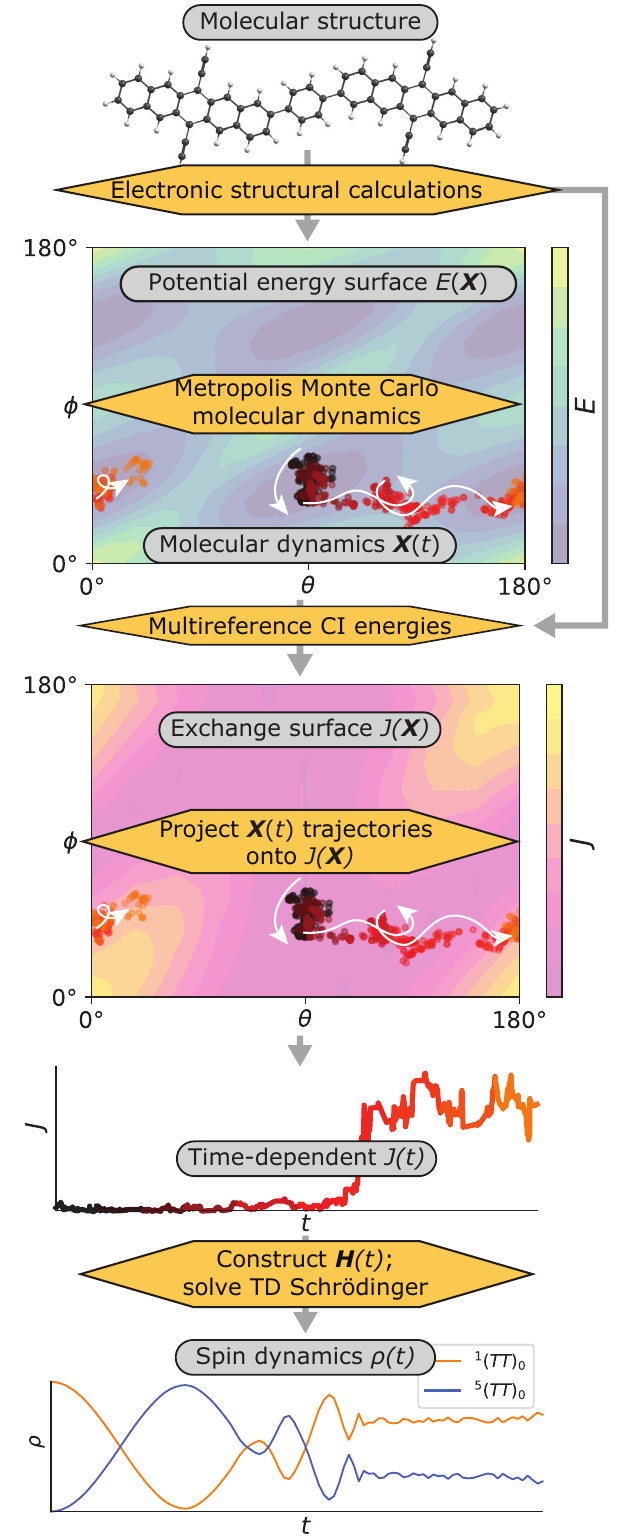}
    \caption{An end-to-end pathway from molecular structure to SF spin dynamics. We model known\cite{Trinh2017} SF material \textbf{BP1} over reduced geometry coordinate $\bm{X}(\theta,\phi)$ (\cref{fig:DoF_BP1}). Processes are indicated with yellow boxes and information products are shown in grey lozenges: as the overall schema is modular, the basic procedures used here could easily be replaced with more sophisticated alternatives.} 
    \label{fig:model_sum}
\end{figure}

By restricting our analysis to a geometry coordinate $\bm{X}(\theta,\phi)$ containing only the pentacene-pentacene dihedral $\theta$ and pentacene-bridge dihedral $\phi$, the complexity of our calculations can be reduced from an intractable (3\textit{N}-6) nuclear degrees of freedom to a computationally feasible 2 dimensions.

The schema of the overall model is shown in \cref{fig:model_sum}. First, we use single-reference electronic structural calculations to obtain a potential energy surface (PES) $E(\bm{X})$. We use this PES as the input for a simple Monte Carlo molecular dynamics simulation, generating time-dependent conformation trajectories $\bm{X}(t)$. We then use multireference configuration interaction calculations in a  (4e,4o) active space\cite{Schmidt2019AFission} to obtain a surface of spin-spin exchange energies, $J(\bm{X})$, and project the previously-obtained molecular dynamics $\bm{X}(t)$ onto this surface to obtain time-dependent exchange trajectories $J(t)$. Finally, the calculated exchange trajectories $J(t)$ are incorporated into the system spin Hamiltonian (\cref{eq:Sham}) to obtain spin dynamics of the system as solutions to the time-dependent Schr\"odinger equation using previously reported methods.\cite{Collins2019FluctuatingFission}

\begin{figure}
    \includegraphics[scale=\pdfscale]{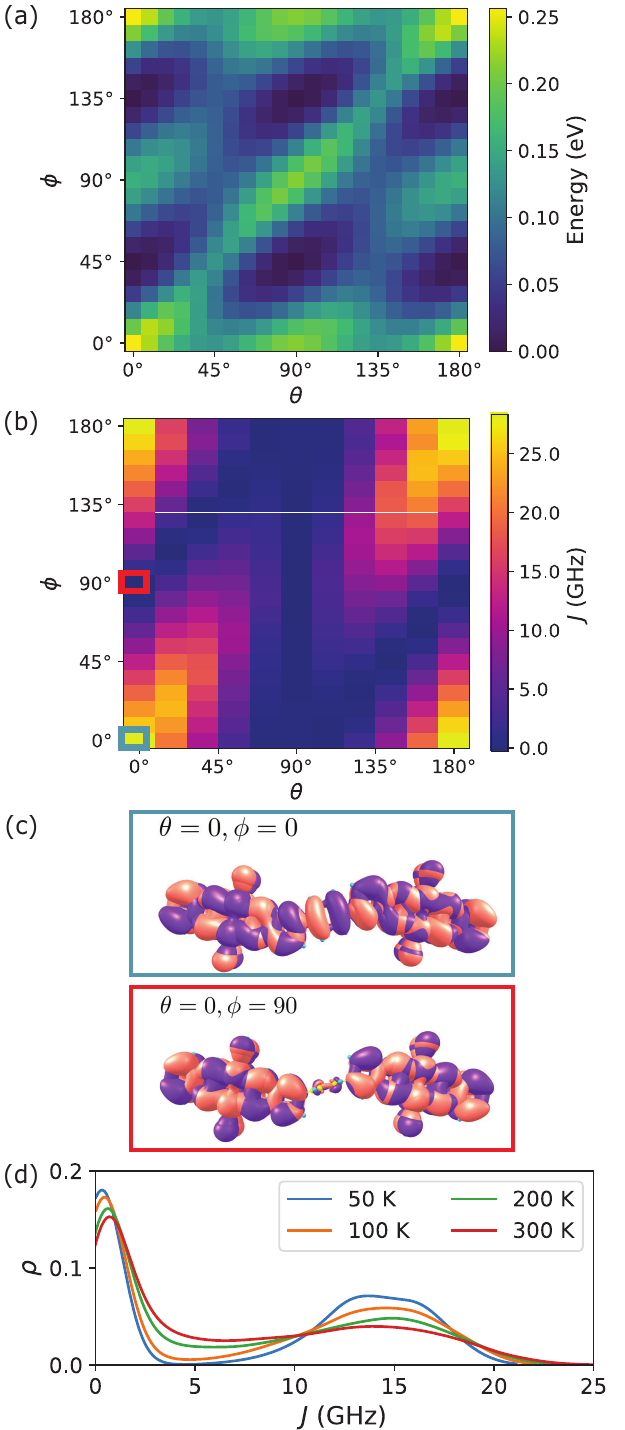}
    \caption{(a) PES for BP1 in the triplet pair $(TT)$ state. (b) The variation of exchange coupling as a function of $\theta$ and $\phi$. Coloured circles indicate where the geometries in (c) lie on the surface. (c) The eight highest energy molecular orbitals with electron density determined by ROHF. The orbitals are rendered for a  co-planar geometry (blue) and an orthogonal geometry (red) to illustrate the difference in orbital overlap across the bridge. (d) Probability distribution of exchange $J$ when \textbf{BP1} is at thermal equilibrium.  }
    \label{fig:Surfaces}
\end{figure}

\section{\label{sec:Results}Results}
\subsection{\label{sec:PES}Potential Energy Surfaces}
The ground state PES of \textbf{BP1} was calculated at the B3LYP~\cite{becke93a}/6-31G(d)~\cite{petersson88,petersson91} level of theory, implemented in Gaussian \cite{Frisch2016GaussianC.01}. The bulky triisopropyl silane (TIPS) functional groups do not contribute significantly to \textbf{BP1}'s electronic character\cite{Kohler2015ElectronicSemiconductors} and were substituted with hydrogen atoms to reduce computational complexity, consistent with previous work.~\cite{pun2018} A relaxed potential energy surface scan over a $21 \times 21$ point grid spanning $180$ degrees was then completed for the constrained degrees of freedom $\theta$ and $\phi$ (\cref{fig:DoF_BP1}). Due to the symmetry of the molecule, the two $180$ degree scans span the unique energy surface of the configuration space. 

The $^{2S+1}(TT)_m$ triplet pair excited state involves a double excitation and cannot be calculated using standard single reference calculations such as DFT or HF theory~\cite{Jensen2007IntroductionChemistry}. In contrast, ground states of a given spin multiplicity are straightforward to calculate. We thus approximated the  $(TT)$ triplet pair state as equivalent to the lowest energy ($\hat{S}=2$) quintet, allowing us to obtain the excited triplet-pair PES of \textbf{BP1} with a single-reference relaxed scan across $\theta$ and $\phi$ at the ROHF/6-31G(d) level, drawing on prior work in this area\cite{Schmidt2019AFission}. \Cref{fig:Surfaces}(a) shows the resultant PES for BP1 in the triplet pair state. Approximating the triplet pair state as the ground-state quintet fails to capture the spin-spin interactions that distinguish $^5(TT)_m$ from the other $^M(TT)_m$ states, but the energy scale of these interactions are $\sim10^5$ smaller than the electronic PES\cite{Tayebjee2017QuintetFission} and are thus unlikely to influence nuclear conformational dynamics. As the ground and triplet pair states were calculated at different levels of theory (B3LYP and ROHF respectively), direct comparison of the resultant energies is not reasonable. Only the comparative energies within the surfaces are relevant to simulate the dynamics of BP1 and are sufficient for our model.

For both singlet and quintet states, the calculated PES consists of four local minima separated by rotational barriers of $0.1$-$0.3\text{ eV}$. One dimensional slices along the $\theta$ and $\phi$ axes resemble double well potentials with minima located at approximately $35\degree$ and $145\degree$. This is comparable to experimentally determined parameters of biphenyl\cite{Imamura1968TheCompounds,Bastiansen1985StructureDerivatives}, a structurally analogous compound, which has an $\approx0.1\text{ eV}$ barrier to rotation about minima located at approximately $\pm35\degree$. In the triplet state, the locations of the double well minima are shifted slightly towards planarity and the rotational barrier decreases in height.

\subsection{Exchange Coupling}
The inter-triplet exchange coupling $J$ was obtained from the energetic separation ($\Delta_{SQ}$) of the singlet and quintet spin states, where $\Delta_{SQ}=3J$ \cite{Kollmar1982TheorySplittings,Benk1981TheoryStructures}. These energies were calculated using a multireference configuration interaction (CI) approach within a (4e,4o) active space, following a methodology previously applied to a model system of ethylene dimers\cite{Schmidt2019AFission}. The CI active space was constructed from the four highest-energy orbitals obtained through ROHF/6-31G(d) calculations of the ground-state. Calculations were implemented in the GAMESS program~\cite{Schmidt1993GeneralSystem}, with all determinants considered. Energies of the near-degenerate $^1(TT)$, $^3(TT)$, and $^5(TT)$ spin states were obtained over the $\bm{X}(\theta,\phi)$ coordinate and used to determine the sign and magnitude of $J$. An example output of this calculation can be found in \cref{sec:SI-Exchange}. 

The coordinate-dependent exchange surface of \textbf{BP1}, $J(\bm{X})$, is shown in \Cref{fig:Surfaces}~(b). Consistent with expectations, $J\simeq0$ wherever at least one chromophore is orthogonal to the phenyl bridge ($[\phi-\theta]=90\degree,270\degree,\dots$) or where both chromophores are orthogonal ($\theta = 90\degree,270\degree,\dots$). The sign of $J$ is almost always positive ($J\geq0$), meaning that the system shows antiferromagnetic coupling with $^5(TT)_0$ at a
higher energy than the $^1(TT)_0$ singlet which differs from some previous reports\cite{Rugg2022a,Hasobe2022}. However, there are rare geometries where orthogonal ring orientations result in no orbital overlap and weakly ferromagnetic spin-spin coupling---i.e., where $J\lesssim0$ (\cref{fig:Surfaces} (c)).

For pentacene dimers such as \textbf{BP1}, the zero field splitting parameter $D$ is on the order of $1$ GHz~\cite{Tayebjee2017QuintetFission}. As the calculated exchange coupling $J(\bm{X})$ ranges from $J\sim 0$ GHz to $J\sim30$ GHz,  \textbf{BP1} is able to satisfy the requirement that conformational dynamics must access $J(t)$ values that are sometimes small and sometimes large relative to $D$. This supports the argument that low-energy dynamics on a simple intramolecular coordinate  $\bm{X}$ are sufficient to drive spin state mixing \emph{via} dynamic  exchange fluctuations.

The time-averaged occupancy of the exchange surface is determined by thermal Boltzmann statistics of the electronic PES, with $J(t)$ primarily sampling $J(\bm{X})$ values corresponding to local minima of $E(\bm{X})$.  
Figure \ref{fig:Surfaces} (d) shows probability distributions of $J$ weighted by Boltzmann statistics of the underlying $E(\bm{X})$ PES. We find that conformational energy minima $E(\bm{X})$ correspond to $J(\bm{X})$ regions of both large and small $J$ relative to $D$ (\cref{fig:Surfaces} (b)). Intermediate values of $J$ are less populated, supporting the validity of recent work\cite{Collins2022QuintetFission} modelling trajectories through this space as hops between large and small $J$. As temperature increases, the distributions flatten and these intermediate values become more accessible to the molecule although a small bias towards low values of $J$ is retained. 

\subsection{Metropolis Monte Carlo Simulations}
\label{sec:MC sims}
\subsubsection{Dynamics of unrestricted \textbf{BP1}}
\begin{figure*}
    \includegraphics[scale=\pdfscale]{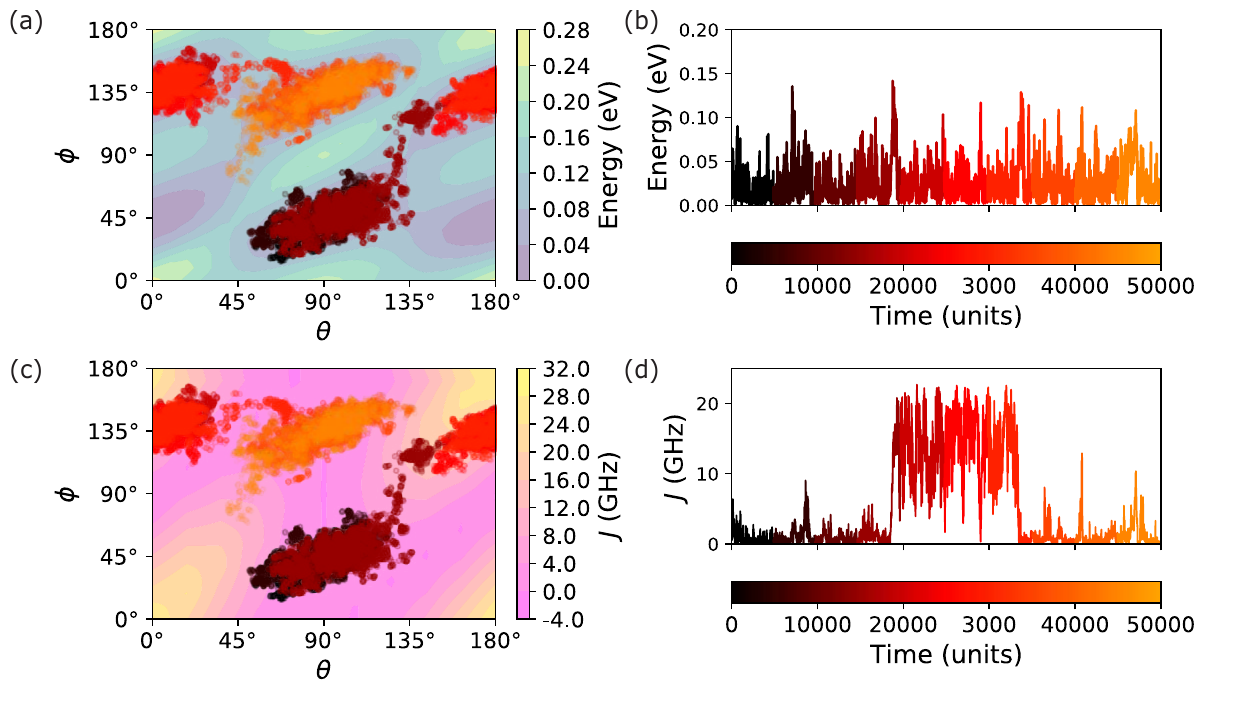}
    \caption{(a) Simulated molecular dynamics $\bm{X}(t)$ obtained by metropolis Monte Carlo evolution over the quintet PES of \textbf{BP1} at 200 K. (b) Long periods of local exploration within an energy minimum are broken up by infrequent site-hopping over higher-energy saddle points to access adjacent locally-stable conformations. (c) $\bm{X}(t)$ mapped onto the calculated exchange surface $J(\bm{X})$. (d) Transitions between local energy minima produce a noisy step function in exchange space. The colour map indicates the progression of time from the start of the simulation (black) to the end of the simulation (yellow).}
    \label{fig:MC trajectory}
\end{figure*}
The molecular dynamics of \textbf{BP1} over geometry coordinate $\bm{X}(\theta,\phi)$ were simulated using a Metropolis Monte Carlo algorithm. Simulations were completed as follows (see \cref{sec:SI-MC} for details): (i) The molecule was initialised in a conformation sampled randomly from the Boltzmann-weighted ground state PES occupancy. (ii) The two dimensional PES of the quintet state (as a surrogate for $(TT)$) was then sampled using a Metropolis-Hastings algorithm modified from prior work\cite{Tiana2007UseChains}. (iii) The trajectories produced by the algorithm were mapped onto the exchange surface calculated by CI and used to generate a time dependent value for exchange $J(t)$. %

\Cref{fig:MC trajectory} shows a typical trajectory for \textbf{BP1} at 200 K, demonstrating the ability of this model to capture both low-energy fluctuations around a conformational energy minimum and higher-energy hops between adjacent minima. These molecular dynamics trajectories $\bm{X}(t)$ can then be projected onto the exchange surface $J(\bm{X})$ (\cref{fig:MC trajectory} (c)) to obtain a trajectory of dynamics-induced fluctuating exchange, $J(t)$ (d). The exchange fluctuations calculated by this approach appear as noisy step functions: the steps between $J\approx0$ and $J\approx15~\mathrm{GHz}$ result from rapid transient passage between different locally-stable conformations, while the superimposed noise comes from low-energy dynamics within a conformational energy minimum.


The form of $J(t)$ was found to be robust to changes in the model parameters. The length scale of steps and the maximum individual step-size impact the rate at which transitions occur and the variance in exchange space. However, the core behaviour of a step function between two regimes of $J$ is reproduced. 

It is important to note the gas phase PESs calculated in \cref{sec:PES} are not ideal analogs for molecules in the condensed phases used for experimental studies of singlet fission (molecular solvents for room-temperature optical measurements, or frozen glasses for low-temperature spin resonance). Nevertheless, these energy landscapes provide a straightforward proof of concept that can be substituted by more accurate models in future iterations of this analytic schema.
%

\subsubsection{Restricting large-amplitude chromophore dynamics}
\label{sec:constrained}
Experimental spin resonance measurements of SF molecular dimers typically  occur in frozen solution at cryogenic temperatures or with the molecules held in solid matrices~\cite{Tayebjee2017QuintetFission,Sakai2018MultiexcitonPairs,Pun2019Ultra-fastDesign}. Whether large changes in geometry can still occur in these systems is not settled~\cite{Kobori2020GeometriesFissions}. While the local heating effect of high-energy laser photoexcitation may temporarily allow thermal access to molecular dynamics, large-amplitude dynamics involving bulky chromophore rotations seem unlikely under these circumstances.\cite{Concistre2014FreezingNMR}. 
We thus consider the case of restricted molecular dynamics in which the local environment prevents inter-chromophore rotations of the bulky pentacene groups, freezing $\theta$  such that all dynamics occur on the chromophore-bridge coordinate $\phi$.\cite{Illig2016ReducingMotions, Kwon2015SuppressingMaterials,Nobukawa2013DynamicsTransition}. Modelling the impact of restricted  motion on SF spin dynamics may help inform the types of geometry changes required for SF.

\begin{figure}
    \includegraphics[scale=\pdfscale]{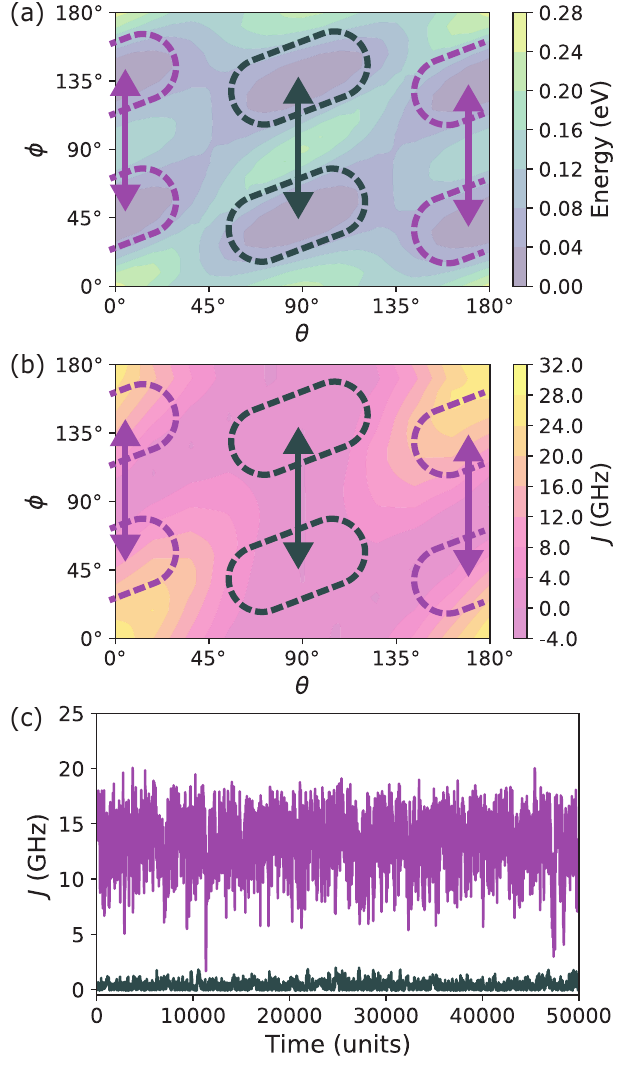}
    \caption{Effect of freezing chromophore-chromophore dynamics $(\theta)$ and only allowing bridge rotations $(\phi)$, as might occur in a solid matrix. (a) Conformational dynamics between locally-stable geometries are still possible, but (b) no longer result in large $J$ changes. (c) Instead, $J(t)$ fluctuates within  high-exchange or low-exchange regimes (purple and grey lines, respectively) showing thermal noise without step changes.}
    \label{fig:constrained surfaces}
\end{figure}

Restricting the $\theta$ degree of freedom only allows transitions between conformations that are accessible via bridge rotations ($\phi$). \Cref{fig:constrained surfaces} depicts these allowed transitions in energy and exchange space, showing that the previously-observed step changes in $J$ do not occur in the absence of inter-chromophore rotations. 
Instead, the trajectories in exchange space are simply thermal fluctuations within the fixed high- or low-$J$ conformation as defined by the initial interchromopore angle $\theta$. In the low exchange regime, $J\sim0$ GHz, spin states will be able to mix but will never form long-lived states with well defined spin such as those observed in nutation experiments~\cite{Tayebjee2017QuintetFission}. In contrast, if the molecule is in the large exchange regime $J\sim15$ GHz, coherent mixing between the singlet and quintet spin states is prohibited. Well-defined $(TT)$ spin states with $S\neq0$ may still be accessed through other pathways such as stochastic spin-mixing in a high-$J$ manifold\cite{Collins2022QuintetFission}, but these results suggest that acene dimers with a single (1,4-phenylene) bridging motif may have limited pathways for accessing high-spin $(TT)$ states when in restricted environments. Molecules with additional degrees of freedom and multiple conformational minima, such as poly(1,4)-phenylene bridges,\cite{Tayebjee2017QuintetFission} are more likely to enable variation between high- and low-$J$ in restricted geometries. 

\subsection{\label{sec:Spin Ds}Spin Dynamics}
The initial product of SF is a pair of triplet excitons localised on coupled but separate molecular sites. Here we treat the spin Hamiltonian of the triplet pair following the methodology outlined previously\cite{Collins2019FluctuatingFission,Collins2022QuintetFission}. The Hamiltonian can be written as the sum of the Zeeman interaction ($\hat{H}_\textrm{Z}$) from an externally applied magnetic field ($\bm{B}$), the zero-field splitting interactions ($\hat{H}_{\textrm{zfs}}$), and a time-dependent intertriplet exchange interaction ($\hat{H}_{\textrm{ee}}$)
\begin{subequations}
\begin{eqnarray}
    \hat{H} & = & \hat{H}_\textrm{Z} + \hat{H}_{\textrm{zfs}} +\hat{H}_{\textrm{ee}},\\\nonumber\\
    \hat{H}_\textrm{Z} & = & \mu_\textrm{B} g\sum_{i=1,2}\sum_{j=x,y,z} B_j\hat{S}_{ij},\\\nonumber\\
    \hat{H}_{\textrm{zfs}} & = & \sum_{i=1,2} \hat{S}_i \cdot \bm{D}_i \cdot \hat{S}_i, \\\nonumber\\
    \hat{H}_{\textrm{ee}} & = & J(t)\hat{S}_1\cdot\hat{S}_2,
\end{eqnarray}
\end{subequations}
where $\mu_\textrm{B}$ is the Bohr magneton, $g$ is the Landé \textit{g-factor}, and $\hat{S}_i$ represents the spin operator for the triplet pair on the $i$th chromophore. $\bm{D}_i$ is the ZFS tensor of the $i$th chromophore, which is often described by axiality and rhombicity parameters $D$ and $E$ along with an orientation term.

In this work, we apply our simulated Monte Carlo trajectories to obtain a realistic form for $J(t)$ and additionally, incorporate a time dependent ZFS interaction into the model. $J(t)$ is determined directly by the method described above. The time-dependent ZFS interaction was modelled by defining a time-dependent rotation matrix $R_{\textrm{rel}}(\Theta_{1,2})$ that describes the relative inter-triplet orientation of the chromophores over the course of the trajectory. The relative orientation was parameterised by three Euler angles in the $zyz$ convention, $\Theta_{1,2}=(\theta_z,\theta_y,\theta_z')$.
\begin{eqnarray}
\bm{D}_2&=& R_{\textrm{rel}}\bm{D}_1 R_{\textrm{rel}}^T,\\
R_{\textrm{rel}} &=& R_z(\theta_z') R_y(\theta_y) R_z(\theta_z),
\end{eqnarray}
This orientation was approximated by the chromophore-chromophore dihedral $\theta$ over time. As $\theta$ varies, $\bm{D}_1$ remains fixed along the $z$-axis while $\bm{D}_2$ rotates about the $y$-axis as a function of $R_{\textrm{rel}}(t)=R_{\textrm{rel}}(0,\theta(t),0)$. The parameters of our spin Hamiltonian for \textbf{BP1} are summarised in \cref{tab:spin params}. The time-dependent Schrödinger equation for this system was then solved using the \verb+mesolve+ function in Python quantum toolbox software QuTip~\cite{Johansson2012QuTiP:Systems}.
\begin{table}
\caption{\label{tab:spin params}Parameters used to model the spin Hamiltonian of \textbf{BP1}. Fixed parameters ($g,D,E$ and $\bm{B})$ are from literature.\cite{Tayebjee2017QuintetFission}}
\begin{ruledtabular}
\begin{tabular}{ll}
Parameter & Value\\
\hline
${g_x=g_y=g_z}$   &  2.002\\[0.2cm]
${D}$ & 1138 MHz\\[0.2cm]
${E}$   & 19 MHz\\[0.2cm]
$\boldsymbol{B}=(B_x,B_y,B_z)$   &   (0,0,350mT)\\[0.2cm]
$J(t)$ & $J(\bm{X}(t))$\\[0.2cm]
$\bm{D}_i$ & $R_{\textrm{rel}}\bm{D}_jR_{\textrm{rel}}^T$\\[0.2cm]
$R_{\textrm{rel}}(t)$ & $R_{\textrm{rel}}\Bigl(0,\theta(t),0\Bigr)$
\end{tabular}
\end{ruledtabular}
\end{table}

Simulated molecular dynamics trajectories take place over a period of Monte Carlo `time', $t_\textrm{MC}$, related to the number of steps taken during the simulation. This  $t_\textrm{MC}$ time unit is purely a function of the numerical simulation and does not possess physical significance. To model $J(t)$ in physically significant time units, $t_\textrm{MC}$ was scaled to a period of real time $t=\alpha t_\textrm{MC}$ where $\alpha$ is the timescale parameter. To determine realistic values of $\alpha$, upper and lower bounds for minima transition rates were estimated. With $D\simeq1$GHz, the oscillation between the $^1(TT)_0$ and $^5(TT)_0$ states when mixed by the ZFS interaction has period $\sim1$ ns.\cite{Collins2019FluctuatingFission} If transitions between $J$ regimes occur much faster than $D$, these processes become decoupled and stochastic $J$-fluctuations are no longer effective at mixing different spin states. Quintets have been observed with coherence times on the order of $\sim100$ ns at temperatures below $100$ K~\cite{Weiss2017StronglySemiconductor,Tayebjee2017QuintetFission}. By assuming quintets rapidly decohere after transitioning to $J\sim0$ GHz, time spent in the large exchange regime can be considered an upper bound for coherence times. Hence, the average $t_\textrm{MC}$ spent in large exchange minima before transitioning was compared and scaled to experimental values of $D$ and coherence times to give a realistic range for $\alpha$ ($\alpha=0.5$ ps to $50$ ps). This approach to scaling Monte Carlo time is sufficient to demonstrate the viability of quintet formation by conformation-induced exchange fluctuations, but cannot provide \emph{ab initio} predictions of dynamics without scaling. A logical next step beyond the scope of this work would be to replace or re-scale the Monte Carlo trajectories used here with results from molecular dynamics simulations.

\subsubsection{Quintet Formation \textit{Via} Simulated Trajectories}
\begin{figure}
    \includegraphics[scale=\pdfscale]{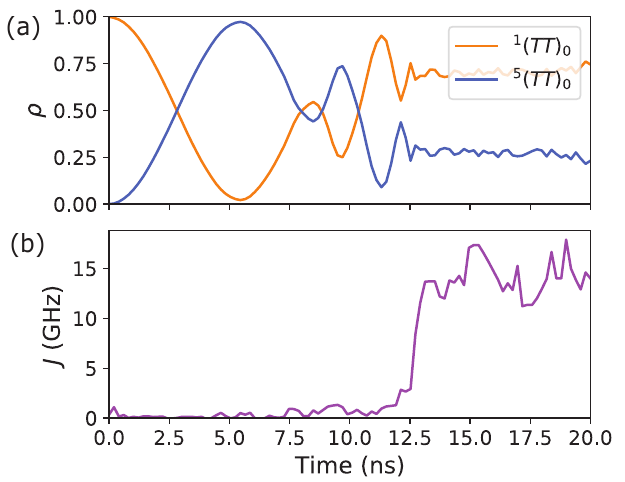}
    \caption{(a) $^1(TT)_0$ and $^5(TT)_0$ projections of the system wavefunction $\psi(t)$ evolving under (b) modelled time-dependent exchange coupling. Initially, the singlet and quintet states mix while $J\simeq0$. Then, when $J$ quickly becomes too large for the spin states to be mixed by $\hat{H}_{zfs}$, the probability density stabilises.}
    \label{fig:simple spin evo}
\end{figure}

As in \cref{fig:MC trajectory}, $J(t)$ dynamics for the unrestricted \textbf{BP1} dimer appear as a noisy step function between regimes of high and low $J$. \Cref{fig:simple spin evo} shows a numerical solution to the time-dependent Schrödinger equation for a sample $J(t)$ trajectory with $\alpha=3\times10^{-2}$ ns, demonstrating that these dynamics can lead to well-resolved $^5(TT)$ quintet multiexcitons. In this trajectory, \textbf{BP1} is initially in a low-$J$ conformation where $^5(TT)_0$ and  $^1(TT)_0$ mix coherently. The system then jumps to a new conformation with $J\approx15$ GHz, resolving the spin states in energy and freezing mixing with a significant expected population of the $\hat{S}=2$ quintet. Having demonstrated the ability of this model to reproduce quintet formation in a single molecule, we now move to consider the population-averaged spin dynamics of an ensemble of molecules evolving according to this model. 

\subsubsection{Population Dynamics}

\begin{figure*}
    \includegraphics[scale=\pdfscale]{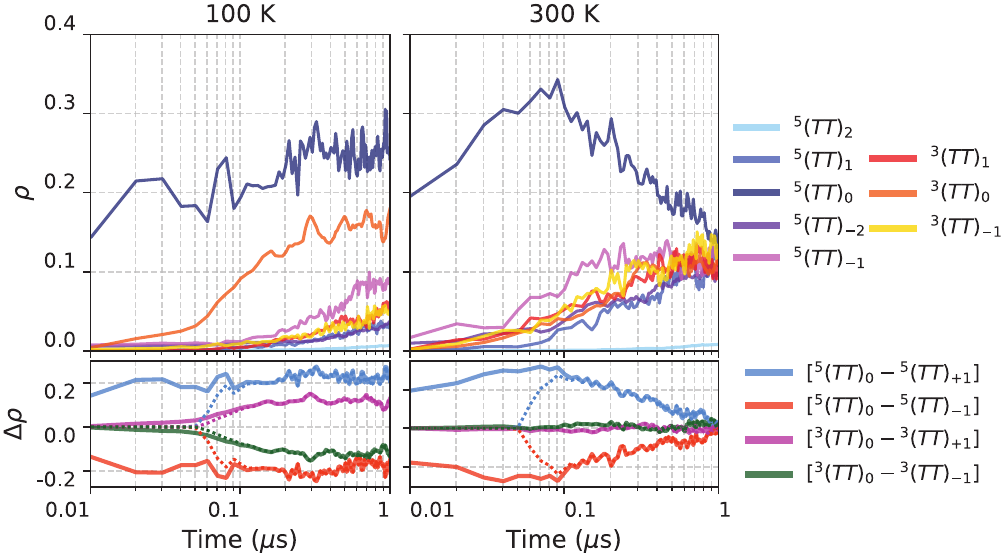}
    \caption{The spin dynamics of an ensemble of 50 \textbf{BP1} molecules over time at $100$ and $300$ Kelvin. Upper plots depict the population densities while lower plots display the difference in spin populations (spin polarisation). Dotted lines imitate the signal observed by time resolved ESR by filtering the densities with an exponential envelope that rises in at $t=50$ ns to imitate the response time of the experimental apparatus. Rapid formation of the $^5(TT)_0$ quintet occurs for all temperatures with the formation of the $^3(TT)_0$ triplet strongly temperature dependent. The spin densities converge as mixing allows all states other than  $^5(TT)_2$ to be populated.}
    \label{fig:spin pops}
\end{figure*}
Ensemble simulations of $50$ molecules with motion scaled to $\mathrm{1~\mu{}s}$ (details in \cref{sec:SI-Spin calcs}) show spin dynamics  qualitatively consistent with experimental observations~\cite{Tayebjee2017QuintetFission,Weiss2017StronglySemiconductor}(\cref{fig:spin pops}). Quintets are populated within $\sim1$ ns by molecules initialised in $J\sim0$ GHz minima, allowing immediate spin mixing. At later times, $m_s \neq0$ quintets form more slowly \textit{via} $\hat{H}_{zfs}$-mediated mixing. As $J_{iso}(t)\geq 0$ and $B>0$, the $m_s > 0$ $^5(TT)_m$ sublevels remain separated from $^1(TT)$ by the Zeeman interaction and a bias is seen towards the formation of $m_s \leq 0$ spin states. Here the $m_s = -1$ quintet is formed preferentially as $J$ is less frequently on resonance with the $^1(TT)_0{}\leftrightarrow{}^5(TT)_{-2}$ crossing. These conditions of $B>0$ and $J>0$ also mean that the $m_s > 0$ quintet states do not easily mix with $^1(TT)_0$ and form slowly. Additionally, the mixing of the triplet with singlet and quintet states is limited by the symmetry of the chromophores~\cite{Bayliss2016SpinFission} and the rate of formation is slow.  At low temperatures, where molecules are likely to stay confined to local minima for long periods, the triplet state $^3(TT)_0$ forms with significant polarization from the rest of the triplet manifold. 

The formation of polarised spin states is necessarily a transient, kinetic phenomenon: over very long time periods, all spin states are populated and tend towards a near-equal Boltzmann population distribution. Increasing the temperature of the system decreases the time spent in a single local energy minimum and increases the frequency of exchange variations. This inhibits the slow mixing of the $^3(TT)_0$ triplet and increases the number of spin transitions that see a resonant frequency of $J$, and we see less spin polarisation and rapid decay of the initial $^5(TT)_0$ excess at $300$K compared to $100$K. Transitions from the high-$J$ local energy minima to the $J\sim0$ GHz global minima (\cref{fig:Surfaces}(d)) also occur more rapidly at high temperatures, leading to a greater population of $^5(TT)_0$ at early times. 
\section{Discussion}
\label{sec:discussion}
\subsection{Comparison to Experimental Findings}
The results above show that conformation-induced exchange fluctuations in molecular SF dimers can drive spin dynamics and quintet formation. In addition, our simulations suggest that the $^5(TT)_{\pm1,2}$ and $^3(TT)_{0,\pm1}$ states form with significant initial spin polarisation. The polarisations of these spin states can be probed experimentally with transient electron spin resonance (ESR) spectroscopy\cite{Tayebjee2017QuintetFission,Weiss2017StronglySemiconductor}, which measures the $B$-dependent absorption and/or emission of microwaves corresponding to $\Delta m=\pm1$ spin-spin transitions. Where overlapping transitions exist, resolving the spectra and interpreting the signal can be complex due to the overlap of multiple transitions occurring at the same field. Most relevant to these works, the $^5(TT)_{\pm1\leftrightarrow2}$ and $^3(TT)_{0\leftrightarrow\pm1}$ transitions overlap with those of free triplets ($^3T_{0\leftrightarrow\pm1}$). meaning that the delayed cw-ESR signal previously attributed to free triplets\cite{Tayebjee2017QuintetFission,Weiss2017StronglySemiconductor} may also have contributions from these spin states. Very recently, $^5(TT)_{\pm1\leftrightarrow2}$ transitions have been experimentally observed for the first time via coherent pulsed ESR (p-ESR)\cite{macdonald_tayebjee_collins_kumarasamy_sanders_sfeir_campos_mccamey_2023}, confirming that SF spin dynamics are more complex than first thought. 

\subsection{Comparison to Theoretical Models}

Proposed models for exchange fluctuation include nuclear reorganisation from small to large exchange regimes\cite{Collins2019FluctuatingFission}, stochastic switching between two stable regimes (e.g a double potential energy well)\cite{Kobori2020GeometriesFissions,Collins2022QuintetFission} and harmonic oscillations within a varying exchange regime\cite{Collins2022QuintetFission}. The quintets formed can be categorised as fast (those that form coherently via ZFS mixing at $J\sim0$\cite{Collins2019FluctuatingFission}), and slow (those that form via stochastic exchange fluctuations at $J>>D$\cite{Collins2022QuintetFission}). The model we present here can inform the validity of each of these theoretical approaches and guide parameter choice for their implementation.

The formation of a fast population of quintets \textit{via} mixing in a $J\sim0$ conformation before transitioning to $J>>D$\cite{Collins2019FluctuatingFission} is consistent with the PES, exchange surface and simulated dynamics of \textbf{BP1}. These transitions are expected to occur rapidly relative to the time spent in each regime given the high energy of intermediate geometries. Our results are also consistent with a kinetic switching model driven by minima hopping. For these types of exchange fluctuations to occur, it is predicted that large-scale chromophore motions may be required (\cref{sec:constrained}).

Recent work has demonstrated that (slow) quintets can be formed when the dimer is confined to a large $J$ regime. Collins et al.\cite{Collins2022QuintetFission} found that diabatic $^1(TT)_0{}\leftrightarrow{}^5(TT)_{0}$ transitions can be driven by noise in the exchange signal. When our \textbf{BP1} model restricts large-scale chromophore motions, a noisy, high $J$ signal, consistent with this approach is observed (\cref{fig:constrained surfaces} (c)). 

Based on this work, we can identify two qualitatively different dynamic regimes, which are likely to lead to two different quintet ensembles, consistent with theoretical models\cite{Collins2019FluctuatingFission,Collins2022QuintetFission}. The first ensemble, driven by relatively slow dynamical transitions between high and low exchange, forms rapidly ($\sim$ ns) and has short coherence times ($<<\mu$s) whilst the other, fluctuations within a high exchange geometric minima, can lead to stochastic generation of quintets~\cite{Collins2022QuintetFission} which take longer to form ($\sim\mu$s), and have longer coherence times ($\sim\mu$s). Relaxation to a $J\sim0$ minima will allow quintets to decohere by ZFS mixing. As such, the time between minima transitions may provide an upper bound on the coherence time of quintets. Similarly, as the molecule must transition to a $J>>D$ minima for non-stationary quintets to become well defined, the hopping rate places a lower bound on formation time. From this, we see that our BP1 model is consistent either with fast quintets (if minima hops occur with frequency $f\sim$ GHz) or with slow quintets ($f<$ MHz), but not both. Calibrating the timescale of our simulation with short molecular dynamics calculations should help determine which of these mechanisms dominate. The occurrence of both mechanisms simultaneously may be explained by more complex PESs for dimers with additional bridge degrees of freedom. This may allow some dimers to `trapped' within the large $J$ regime while others have the ability to transition between regimes easily. A mixture of molecules in which some have large-scale chromophore motions restricted and/or enabled by solvent interactions may also arise.
\subsection{Implications for Molecular Design}
Molecular design is increasingly used to enhance the properties of SF materials\cite{Pun2019Ultra-fastDesign,Kobori2020GeometriesFissions,Jacobberger2022UsingCrystals}. A computational pipeline from molecular structure through to spin dynamics is likely to prove useful as a test-bed for such design. Our results support experimental evidence\cite{Nakamura2021SynergeticFission} that more flexible dimers allow quintet states to be formed more rapidly. Rapid population of the quintet manifold prevents decay of the multiexcition \textit{via} triplet-triplet annihilation and may be beneficial for the quantum yield of SF photovoltaic devices\cite{Chen2019Quintet-tripletTemperature}. Decay into low $J$ minima allows quintet spin states to relax and may limit coherence times. Rigid dimers may produce slowly formed, long-lived coherent states that will be beneficial for quantum information applications. The ability to propagate molecular structure data through to simulated spin dynamics could allow novel designs to be developed and suitability assessed.
\section{Conclusion}
We have developed a modular computational pipeline from molecular structure data to physically informed spin dynamics. The model affirms recent observations that the dynamics of the $^1(TT)_0$ state are more complex than models which proceed in a step-wise fashion through the spin manifolds might suggest. The links between the conformational dynamics of the host molecule and evolution of spin states is an important consideration in understanding singlet fission in sufficient detail to design new materials for technological functions. We also show that  a number of spin states may contribute to ESR signals previously attributed to free triplets. This methodology informs the validity of existing theoretical approximations for $J(t)$. It's modular design means it can readily be adapted to other approaches and help determine physically sensible spin simulation parameters. The ability to probe the impact of molecular structure changes on spin dynamics means this pipeline has the potential to form a test bed for molecular design.

\begin{acknowledgments}
This work is supported by the Australian Research Council via the ARC Centre of Excellence in Exciton Science (CE170100026). This work was conducted using the National Computational Infrastructure (NCI), which is supported by the Australian Government. 
\end{acknowledgments}


\begin{thebibliography}{52}%
	\makeatletter
	\providecommand \@ifxundefined [1]{%
		\@ifx{#1\undefined}
	}%
	\providecommand \@ifnum [1]{%
		\ifnum #1\expandafter \@firstoftwo
		\else \expandafter \@secondoftwo
		\fi
	}%
	\providecommand \@ifx [1]{%
		\ifx #1\expandafter \@firstoftwo
		\else \expandafter \@secondoftwo
		\fi
	}%
	\providecommand \natexlab [1]{#1}%
	\providecommand \enquote  [1]{``#1''}%
	\providecommand \bibnamefont  [1]{#1}%
	\providecommand \bibfnamefont [1]{#1}%
	\providecommand \citenamefont [1]{#1}%
	\providecommand \href@noop [0]{\@secondoftwo}%
	\providecommand \href [0]{\begingroup \@sanitize@url \@href}%
	\providecommand \@href[1]{\@@startlink{#1}\@@href}%
	\providecommand \@@href[1]{\endgroup#1\@@endlink}%
	\providecommand \@sanitize@url [0]{\catcode `\\12\catcode `\$12\catcode
		`\&12\catcode `\#12\catcode `\^12\catcode `\_12\catcode `\%12\relax}%
	\providecommand \@@startlink[1]{}%
	\providecommand \@@endlink[0]{}%
	\providecommand \url  [0]{\begingroup\@sanitize@url \@url }%
	\providecommand \@url [1]{\endgroup\@href {#1}{\urlprefix }}%
	\providecommand \urlprefix  [0]{URL }%
	\providecommand \Eprint [0]{\href }%
	\providecommand \doibase [0]{http://dx.doi.org/}%
	\providecommand \selectlanguage [0]{\@gobble}%
	\providecommand \bibinfo  [0]{\@secondoftwo}%
	\providecommand \bibfield  [0]{\@secondoftwo}%
	\providecommand \translation [1]{[#1]}%
	\providecommand \BibitemOpen [0]{}%
	\providecommand \bibitemStop [0]{}%
	\providecommand \bibitemNoStop [0]{.\EOS\space}%
	\providecommand \EOS [0]{\spacefactor3000\relax}%
	\providecommand \BibitemShut  [1]{\csname bibitem#1\endcsname}%
	\let\auto@bib@innerbib\@empty
	\bibitem [{\citenamefont {Smith}\ and\ \citenamefont
		{Michl}(2010)}]{Smith2010SingletFission}%
	\BibitemOpen
	\bibfield  {author} {\bibinfo {author} {\bibfnamefont {M.~B.}\ \bibnamefont
			{Smith}}\ and\ \bibinfo {author} {\bibfnamefont {J.}~\bibnamefont {Michl}},\
	}\bibfield  {title} {\enquote {\bibinfo {title} {{Singlet Fission}},}\ }\href
	{\doibase 10.1021/cr1002613} {\bibfield  {journal} {\bibinfo  {journal}
			{Chemical Reviews}\ }\textbf {\bibinfo {volume} {110}},\ \bibinfo {pages}
		{6891--6936} (\bibinfo {year} {2010})}\BibitemShut {NoStop}%
	\bibitem [{\citenamefont {Hanna}\ and\ \citenamefont
		{Nozik}(2006)}]{Hanna2006SolarAbsorbers}%
	\BibitemOpen
	\bibfield  {author} {\bibinfo {author} {\bibfnamefont {M.~C.}\ \bibnamefont
			{Hanna}}\ and\ \bibinfo {author} {\bibfnamefont {A.~J.}\ \bibnamefont
			{Nozik}},\ }\bibfield  {title} {\enquote {\bibinfo {title} {{Solar conversion
					efficiency of photovoltaic and photoelectrolysis cells with carrier
					multiplication absorbers}},}\ }\href {\doibase 10.1063/1.2356795} {\bibfield
		{journal} {\bibinfo  {journal} {Journal of Applied Physics}\ }\textbf
		{\bibinfo {volume} {100}},\ \bibinfo {pages} {074510} (\bibinfo {year}
		{2006})}\BibitemShut {NoStop}%
	\bibitem [{\citenamefont {Tayebjee}, \citenamefont {Gray-Weale},\ and\
		\citenamefont {Schmidt}(2012)}]{Tayebjee2012}%
	\BibitemOpen
	\bibfield  {author} {\bibinfo {author} {\bibfnamefont {M.~J.}\ \bibnamefont
			{Tayebjee}}, \bibinfo {author} {\bibfnamefont {A.~A.}\ \bibnamefont
			{Gray-Weale}}, \ and\ \bibinfo {author} {\bibfnamefont {T.~W.}\ \bibnamefont
			{Schmidt}},\ }\bibfield  {title} {\enquote {\bibinfo {title} {{Thermodynamic
					limit of exciton fission solar cell efficiency}},}\ }\href {\doibase
		10.1021/jz301069u} {\bibfield  {journal} {\bibinfo  {journal} {Journal of
				Physical Chemistry Letters}\ }\textbf {\bibinfo {volume} {3}},\ \bibinfo
		{pages} {2749--2754} (\bibinfo {year} {2012})}\BibitemShut {NoStop}%
	\bibitem [{\citenamefont {Tayebjee}\ \emph {et~al.}(2017)\citenamefont
		{Tayebjee}, \citenamefont {Sanders}, \citenamefont {Kumarasamy},
		\citenamefont {Campos}, \citenamefont {Sfeir},\ and\ \citenamefont
		{McCamey}}]{Tayebjee2017QuintetFission}%
	\BibitemOpen
	\bibfield  {author} {\bibinfo {author} {\bibfnamefont {M.~J.}\ \bibnamefont
			{Tayebjee}}, \bibinfo {author} {\bibfnamefont {S.~N.}\ \bibnamefont
			{Sanders}}, \bibinfo {author} {\bibfnamefont {E.}~\bibnamefont {Kumarasamy}},
		\bibinfo {author} {\bibfnamefont {L.~M.}\ \bibnamefont {Campos}}, \bibinfo
		{author} {\bibfnamefont {M.~Y.}\ \bibnamefont {Sfeir}}, \ and\ \bibinfo
		{author} {\bibfnamefont {D.~R.}\ \bibnamefont {McCamey}},\ }\bibfield
	{title} {\enquote {\bibinfo {title} {{Quintet multiexciton dynamics in
					singlet fission}},}\ }\href {\doibase 10.1038/nphys3909} {\bibfield
		{journal} {\bibinfo  {journal} {Nature Physics}\ }\textbf {\bibinfo {volume}
			{13}},\ \bibinfo {pages} {182--188} (\bibinfo {year} {2017})}\BibitemShut
	{NoStop}%
	\bibitem [{\citenamefont {Kawashima}\ \emph {et~al.}(2022)\citenamefont
		{Kawashima}, \citenamefont {Hamachi}, \citenamefont {Yamauchi}, \citenamefont
		{Nishimura}, \citenamefont {Nakashima}, \citenamefont {Fujiwara},
		\citenamefont {Kimizuka}, \citenamefont {Ryu}, \citenamefont {Tamura},
		\citenamefont {Saigo}, \citenamefont {Onda}, \citenamefont {Sato},
		\citenamefont {Kobori}, \citenamefont {Tateishi}, \citenamefont {Uesaka},
		\citenamefont {Watanabe}, \citenamefont {Miyata},\ and\ \citenamefont
		{Yanai}}]{Kawashima2022}%
	\BibitemOpen
	\bibfield  {author} {\bibinfo {author} {\bibfnamefont {Y.}~\bibnamefont
			{Kawashima}}, \bibinfo {author} {\bibfnamefont {T.}~\bibnamefont {Hamachi}},
		\bibinfo {author} {\bibfnamefont {A.}~\bibnamefont {Yamauchi}}, \bibinfo
		{author} {\bibfnamefont {K.}~\bibnamefont {Nishimura}}, \bibinfo {author}
		{\bibfnamefont {Y.}~\bibnamefont {Nakashima}}, \bibinfo {author}
		{\bibfnamefont {S.}~\bibnamefont {Fujiwara}}, \bibinfo {author}
		{\bibfnamefont {N.}~\bibnamefont {Kimizuka}}, \bibinfo {author}
		{\bibfnamefont {T.}~\bibnamefont {Ryu}}, \bibinfo {author} {\bibfnamefont
			{T.}~\bibnamefont {Tamura}}, \bibinfo {author} {\bibfnamefont
			{M.}~\bibnamefont {Saigo}}, \bibinfo {author} {\bibfnamefont
			{K.}~\bibnamefont {Onda}}, \bibinfo {author} {\bibfnamefont {S.}~\bibnamefont
			{Sato}}, \bibinfo {author} {\bibfnamefont {Y.}~\bibnamefont {Kobori}},
		\bibinfo {author} {\bibfnamefont {K.}~\bibnamefont {Tateishi}}, \bibinfo
		{author} {\bibfnamefont {T.}~\bibnamefont {Uesaka}}, \bibinfo {author}
		{\bibfnamefont {G.}~\bibnamefont {Watanabe}}, \bibinfo {author}
		{\bibfnamefont {K.}~\bibnamefont {Miyata}}, \ and\ \bibinfo {author}
		{\bibfnamefont {N.}~\bibnamefont {Yanai}},\ }\bibfield  {title} {\enquote
		{\bibinfo {title} {Singlet fission as a polarized spin generator for
				biological nuclear hyperpolarization},}\ }\href {\doibase
		10.26434/chemrxiv-2022-r4636} {\bibfield  {journal} {\bibinfo  {journal}
			{ChemRxiv}\ } (\bibinfo {year} {2022}),\
		10.26434/chemrxiv-2022-r4636}\BibitemShut {NoStop}%
	\bibitem [{\citenamefont {Jacobberger}\ \emph {et~al.}(2022)\citenamefont
		{Jacobberger}, \citenamefont {Qiu}, \citenamefont {Williams}, \citenamefont
		{Krzyaniak},\ and\ \citenamefont
		{Wasielewski}}]{Jacobberger2022UsingCrystals}%
	\BibitemOpen
	\bibfield  {author} {\bibinfo {author} {\bibfnamefont {R.~M.}\ \bibnamefont
			{Jacobberger}}, \bibinfo {author} {\bibfnamefont {Y.}~\bibnamefont {Qiu}},
		\bibinfo {author} {\bibfnamefont {M.~L.}\ \bibnamefont {Williams}}, \bibinfo
		{author} {\bibfnamefont {M.~D.}\ \bibnamefont {Krzyaniak}}, \ and\ \bibinfo
		{author} {\bibfnamefont {M.~R.}\ \bibnamefont {Wasielewski}},\ }\bibfield
	{title} {\enquote {\bibinfo {title} {{Using Molecular Design to Enhance the
					Coherence Time of Quintet Multiexcitons Generated by Singlet Fission in
					Single Crystals}},}\ }\href {\doibase 10.1021/jacs.1c12414} {\bibfield
		{journal} {\bibinfo  {journal} {Journal of the American Chemical Society}\
		}\textbf {\bibinfo {volume} {144}},\ \bibinfo {pages} {2276--2283} (\bibinfo
		{year} {2022})}\BibitemShut {NoStop}%
	\bibitem [{\citenamefont {Merrifield}, \citenamefont {Avakian},\ and\
		\citenamefont {Groff}(1969)}]{Merrifield1969FissionCrystals}%
	\BibitemOpen
	\bibfield  {author} {\bibinfo {author} {\bibfnamefont {R.~E.}\ \bibnamefont
			{Merrifield}}, \bibinfo {author} {\bibfnamefont {P.}~\bibnamefont {Avakian}},
		\ and\ \bibinfo {author} {\bibfnamefont {R.~P.}\ \bibnamefont {Groff}},\
	}\bibfield  {title} {\enquote {\bibinfo {title} {{Fission of singlet excitons
					into pairs of triplet excitons in tetracene crystals}},}\ }\href {\doibase
		10.1016/0009-2614(69)80144-2} {\bibfield  {journal} {\bibinfo  {journal}
			{Chemical Physics Letters}\ }\textbf {\bibinfo {volume} {3}},\ \bibinfo
		{pages} {386--388} (\bibinfo {year} {1969})}\BibitemShut {NoStop}%
	\bibitem [{\citenamefont
		{Merrifield}(1971)}]{Merrifield1971MagneticInteractions}%
	\BibitemOpen
	\bibfield  {author} {\bibinfo {author} {\bibfnamefont {R.}~\bibnamefont
			{Merrifield}},\ }\bibfield  {title} {\enquote {\bibinfo {title} {{Magnetic
					effects on triplet exciton interactions}},}\ }\href {\doibase
		https://doi.org/10.1351/pac197127030481} {\bibfield  {journal} {\bibinfo
			{journal} {Pure and Applied Chemistry}\ }\textbf {\bibinfo {volume} {27}},\
		\bibinfo {pages} {481--498} (\bibinfo {year} {1971})}\BibitemShut {NoStop}%
	\bibitem [{\citenamefont {He}\ \emph {et~al.}(2022)\citenamefont {He},
		\citenamefont {Parenti}, \citenamefont {Campos},\ and\ \citenamefont
		{Sfeir}}]{He2022}%
	\BibitemOpen
	\bibfield  {author} {\bibinfo {author} {\bibfnamefont {G.}~\bibnamefont
			{He}}, \bibinfo {author} {\bibfnamefont {K.~R.}\ \bibnamefont {Parenti}},
		\bibinfo {author} {\bibfnamefont {L.~M.}\ \bibnamefont {Campos}}, \ and\
		\bibinfo {author} {\bibfnamefont {M.~Y.}\ \bibnamefont {Sfeir}},\ }\bibfield
	{title} {\enquote {\bibinfo {title} {Direct {{Exciton Harvesting}} from a
				{{Bound Triplet Pair}}},}\ }\href {\doibase 10.1002/adma.202203974}
	{\bibfield  {journal} {\bibinfo  {journal} {Advanced Materials}\ ,\ \bibinfo
			{pages} {2203974}} (\bibinfo {year} {2022})}\BibitemShut {NoStop}%
	\bibitem [{\citenamefont {Weiss}\ \emph {et~al.}(2017)\citenamefont {Weiss},
		\citenamefont {Bayliss}, \citenamefont {Kraffert}, \citenamefont {Thorley},
		\citenamefont {Anthony}, \citenamefont {Bittl}, \citenamefont {Friend},
		\citenamefont {Rao}, \citenamefont {Greenham},\ and\ \citenamefont
		{Behrends}}]{Weiss2017StronglySemiconductor}%
	\BibitemOpen
	\bibfield  {author} {\bibinfo {author} {\bibfnamefont {L.~R.}\ \bibnamefont
			{Weiss}}, \bibinfo {author} {\bibfnamefont {S.~L.}\ \bibnamefont {Bayliss}},
		\bibinfo {author} {\bibfnamefont {F.}~\bibnamefont {Kraffert}}, \bibinfo
		{author} {\bibfnamefont {K.~J.}\ \bibnamefont {Thorley}}, \bibinfo {author}
		{\bibfnamefont {J.~E.}\ \bibnamefont {Anthony}}, \bibinfo {author}
		{\bibfnamefont {R.}~\bibnamefont {Bittl}}, \bibinfo {author} {\bibfnamefont
			{R.~H.}\ \bibnamefont {Friend}}, \bibinfo {author} {\bibfnamefont
			{A.}~\bibnamefont {Rao}}, \bibinfo {author} {\bibfnamefont {N.~C.}\
			\bibnamefont {Greenham}}, \ and\ \bibinfo {author} {\bibfnamefont
			{J.}~\bibnamefont {Behrends}},\ }\bibfield  {title} {\enquote {\bibinfo
			{title} {{Strongly exchange-coupled triplet pairs in an organic
					semiconductor}},}\ }\href {\doibase 10.1038/nphys3908} {\bibfield  {journal}
		{\bibinfo  {journal} {Nature Physics}\ }\textbf {\bibinfo {volume} {13}},\
		\bibinfo {pages} {176--181} (\bibinfo {year} {2017})}\BibitemShut {NoStop}%
	\bibitem [{\citenamefont {Chen}\ \emph {et~al.}(2019)\citenamefont {Chen},
		\citenamefont {Krzyaniak}, \citenamefont {Nelson}, \citenamefont {Bae},
		\citenamefont {Harvey}, \citenamefont {Schaller}, \citenamefont {Young},\
		and\ \citenamefont {Wasielewski}}]{Chen2019Quintet-tripletTemperature}%
	\BibitemOpen
	\bibfield  {author} {\bibinfo {author} {\bibfnamefont {M.}~\bibnamefont
			{Chen}}, \bibinfo {author} {\bibfnamefont {M.~D.}\ \bibnamefont {Krzyaniak}},
		\bibinfo {author} {\bibfnamefont {J.~N.}\ \bibnamefont {Nelson}}, \bibinfo
		{author} {\bibfnamefont {Y.~J.}\ \bibnamefont {Bae}}, \bibinfo {author}
		{\bibfnamefont {S.~M.}\ \bibnamefont {Harvey}}, \bibinfo {author}
		{\bibfnamefont {R.~D.}\ \bibnamefont {Schaller}}, \bibinfo {author}
		{\bibfnamefont {R.~M.}\ \bibnamefont {Young}}, \ and\ \bibinfo {author}
		{\bibfnamefont {M.~R.}\ \bibnamefont {Wasielewski}},\ }\bibfield  {title}
	{\enquote {\bibinfo {title} {{Quintet-triplet mixing determines the fate of
					the multiexciton state produced by singlet fission in a terrylenediimide
					dimer at room temperature}},}\ }\href {\doibase 10.1073/pnas.1820932116}
	{\bibfield  {journal} {\bibinfo  {journal} {Proceedings of the National
				Academy of Sciences of the United States of America}\ }\textbf {\bibinfo
			{volume} {116}},\ \bibinfo {pages} {8178--8183} (\bibinfo {year}
		{2019})}\BibitemShut {NoStop}%
	\bibitem [{\citenamefont {Kollmar}\ \emph {et~al.}(1982)\citenamefont
		{Kollmar}, \citenamefont {Sixl}, \citenamefont {Benk}, \citenamefont
		{Denner},\ and\ \citenamefont {Mahler}}]{Kollmar1982TheorySplittings}%
	\BibitemOpen
	\bibfield  {author} {\bibinfo {author} {\bibfnamefont {C.}~\bibnamefont
			{Kollmar}}, \bibinfo {author} {\bibfnamefont {H.}~\bibnamefont {Sixl}},
		\bibinfo {author} {\bibfnamefont {H.}~\bibnamefont {Benk}}, \bibinfo {author}
		{\bibfnamefont {V.}~\bibnamefont {Denner}}, \ and\ \bibinfo {author}
		{\bibfnamefont {G.}~\bibnamefont {Mahler}},\ }\bibfield  {title} {\enquote
		{\bibinfo {title} {{Theory of two coupled triplet states - electrostatic
					energy splittings}},}\ }\href {\doibase 10.1016/0009-2614(82)83139-4}
	{\bibfield  {journal} {\bibinfo  {journal} {Chemical Physics Letters}\
		}\textbf {\bibinfo {volume} {87}},\ \bibinfo {pages} {266--270} (\bibinfo
		{year} {1982})}\BibitemShut {NoStop}%
	\bibitem [{\citenamefont {Collins}, \citenamefont {McCamey},\ and\
		\citenamefont {Tayebjee}(2019)}]{Collins2019FluctuatingFission}%
	\BibitemOpen
	\bibfield  {author} {\bibinfo {author} {\bibfnamefont {M.~I.}\ \bibnamefont
			{Collins}}, \bibinfo {author} {\bibfnamefont {D.~R.}\ \bibnamefont
			{McCamey}}, \ and\ \bibinfo {author} {\bibfnamefont {M.~J.}\ \bibnamefont
			{Tayebjee}},\ }\bibfield  {title} {\enquote {\bibinfo {title} {{Fluctuating
					exchange interactions enable quintet multiexciton formation in singlet
					fission}},}\ }\href {\doibase 10.1063/1.5115816} {\bibfield  {journal}
		{\bibinfo  {journal} {Journal of Chemical Physics}\ }\textbf {\bibinfo
			{volume} {151}},\ \bibinfo {pages} {164104} (\bibinfo {year}
		{2019})}\BibitemShut {NoStop}%
	\bibitem [{\citenamefont {Nagashima}\ \emph {et~al.}(2018)\citenamefont
		{Nagashima}, \citenamefont {Kawaoka}, \citenamefont {Akimoto}, \citenamefont
		{Tachikawa}, \citenamefont {Matsui}, \citenamefont {Ikeda},\ and\
		\citenamefont {Kobori}}]{Nagashima2018}%
	\BibitemOpen
	\bibfield  {author} {\bibinfo {author} {\bibfnamefont {H.}~\bibnamefont
			{Nagashima}}, \bibinfo {author} {\bibfnamefont {S.}~\bibnamefont {Kawaoka}},
		\bibinfo {author} {\bibfnamefont {S.}~\bibnamefont {Akimoto}}, \bibinfo
		{author} {\bibfnamefont {T.}~\bibnamefont {Tachikawa}}, \bibinfo {author}
		{\bibfnamefont {Y.}~\bibnamefont {Matsui}}, \bibinfo {author} {\bibfnamefont
			{H.}~\bibnamefont {Ikeda}}, \ and\ \bibinfo {author} {\bibfnamefont
			{Y.}~\bibnamefont {Kobori}},\ }\bibfield  {title} {\enquote {\bibinfo {title}
			{Singlet-fission-born quintet state: Sublevel selections and trapping by
				multiexciton thermodynamics},}\ }\href {\doibase 10.1021/acs.jpclett.8b02396}
	{\bibfield  {journal} {\bibinfo  {journal} {The Journal of Physical Chemistry
				Letters}\ }\textbf {\bibinfo {volume} {9}},\ \bibinfo {pages} {5855--5861}
		(\bibinfo {year} {2018})},\ \Eprint
	{http://arxiv.org/abs/https://doi.org/10.1021/acs.jpclett.8b02396}
	{https://doi.org/10.1021/acs.jpclett.8b02396} \BibitemShut {NoStop}%
	\bibitem [{\citenamefont {Kobori}\ \emph {et~al.}(2020)\citenamefont {Kobori},
		\citenamefont {Fuki}, \citenamefont {Nakamura},\ and\ \citenamefont
		{Hasobe}}]{Kobori2020GeometriesFissions}%
	\BibitemOpen
	\bibfield  {author} {\bibinfo {author} {\bibfnamefont {Y.}~\bibnamefont
			{Kobori}}, \bibinfo {author} {\bibfnamefont {M.}~\bibnamefont {Fuki}},
		\bibinfo {author} {\bibfnamefont {S.}~\bibnamefont {Nakamura}}, \ and\
		\bibinfo {author} {\bibfnamefont {T.}~\bibnamefont {Hasobe}},\ }\bibfield
	{title} {\enquote {\bibinfo {title} {{Geometries and Terahertz Motions
					Driving Quintet Multiexcitons and Ultimate Triplet-Triplet Dissociations via
					the Intramolecular Singlet Fissions}},}\ }\href {\doibase
		10.1021/acs.jpcb.0c07984} {\bibfield  {journal} {\bibinfo  {journal} {Journal
				of Physical Chemistry B}\ }\textbf {\bibinfo {volume} {124}},\ \bibinfo
		{pages} {9411--9419} (\bibinfo {year} {2020})}\BibitemShut {NoStop}%
	\bibitem [{\citenamefont {Collins}\ \emph {et~al.}(2022)\citenamefont
		{Collins}, \citenamefont {Campaioli}, \citenamefont {Tayebjee}, \citenamefont
		{Cole},\ and\ \citenamefont {McCamey}}]{Collins2022QuintetFission}%
	\BibitemOpen
	\bibfield  {author} {\bibinfo {author} {\bibfnamefont {M.~I.}\ \bibnamefont
			{Collins}}, \bibinfo {author} {\bibfnamefont {F.}~\bibnamefont {Campaioli}},
		\bibinfo {author} {\bibfnamefont {M.~J.~Y.}\ \bibnamefont {Tayebjee}},
		\bibinfo {author} {\bibfnamefont {J.~H.}\ \bibnamefont {Cole}}, \ and\
		\bibinfo {author} {\bibfnamefont {D.~R.}\ \bibnamefont {McCamey}},\
	}\bibfield  {title} {\enquote {\bibinfo {title} {{Quintet formation and
					exchange fluctuations: The role of stochastic resonance in singlet
					fission}},}\ }\href {http://arxiv.org/abs/2206.00816} {\bibfield  {journal}
		{\bibinfo  {journal} {arXiv}\ } (\bibinfo {year} {2022})}\BibitemShut
	{NoStop}%
	\bibitem [{\citenamefont {Trinh}\ \emph {et~al.}(2017)\citenamefont {Trinh},
		\citenamefont {Pinkard}, \citenamefont {Pun}, \citenamefont {Sanders},
		\citenamefont {Kumarasamy}, \citenamefont {Sfeir}, \citenamefont {Campos},
		\citenamefont {Roy},\ and\ \citenamefont {Zhu}}]{Trinh2017}%
	\BibitemOpen
	\bibfield  {author} {\bibinfo {author} {\bibfnamefont {M.~T.}\ \bibnamefont
			{Trinh}}, \bibinfo {author} {\bibfnamefont {A.}~\bibnamefont {Pinkard}},
		\bibinfo {author} {\bibfnamefont {A.~B.}\ \bibnamefont {Pun}}, \bibinfo
		{author} {\bibfnamefont {S.~N.}\ \bibnamefont {Sanders}}, \bibinfo {author}
		{\bibfnamefont {E.}~\bibnamefont {Kumarasamy}}, \bibinfo {author}
		{\bibfnamefont {M.~Y.}\ \bibnamefont {Sfeir}}, \bibinfo {author}
		{\bibfnamefont {L.~M.}\ \bibnamefont {Campos}}, \bibinfo {author}
		{\bibfnamefont {X.}~\bibnamefont {Roy}}, \ and\ \bibinfo {author}
		{\bibfnamefont {X.-Y.}\ \bibnamefont {Zhu}},\ }\bibfield  {title} {\enquote
		{\bibinfo {title} {Distinct properties of the triplet pair state from singlet
				fission},}\ }\href {\doibase 10.1126/sciadv.1700241} {\bibfield  {journal}
		{\bibinfo  {journal} {Sci. Adv.}\ }\textbf {\bibinfo {volume} {3}},\ \bibinfo
		{pages} {e1700241} (\bibinfo {year} {2017})}\BibitemShut {NoStop}%
	\bibitem [{\citenamefont {Ringström}\ \emph {et~al.}(2022)\citenamefont
		{Ringström}, \citenamefont {Edhborg}, \citenamefont {Schroeder},
		\citenamefont {Chen}, \citenamefont {Ferguson}, \citenamefont {Tykwinski},\
		and\ \citenamefont {Albinsson}}]{Ringstrom2022}%
	\BibitemOpen
	\bibfield  {author} {\bibinfo {author} {\bibfnamefont {R.}~\bibnamefont
			{Ringström}}, \bibinfo {author} {\bibfnamefont {F.}~\bibnamefont {Edhborg}},
		\bibinfo {author} {\bibfnamefont {Z.~W.}\ \bibnamefont {Schroeder}}, \bibinfo
		{author} {\bibfnamefont {L.}~\bibnamefont {Chen}}, \bibinfo {author}
		{\bibfnamefont {M.~J.}\ \bibnamefont {Ferguson}}, \bibinfo {author}
		{\bibfnamefont {R.~R.}\ \bibnamefont {Tykwinski}}, \ and\ \bibinfo {author}
		{\bibfnamefont {B.}~\bibnamefont {Albinsson}},\ }\bibfield  {title} {\enquote
		{\bibinfo {title} {Molecular rotational conformation controls the rate of
				singlet fission and triplet decay in pentacene dimers},}\ }\href {\doibase
		10.1039/D1SC06285A} {\bibfield  {journal} {\bibinfo  {journal} {Chem. Sci.}\
		}\textbf {\bibinfo {volume} {13}},\ \bibinfo {pages} {4944--4954} (\bibinfo
		{year} {2022})}\BibitemShut {NoStop}%
	\bibitem [{\citenamefont {Schmidt}(2019)}]{Schmidt2019AFission}%
	\BibitemOpen
	\bibfield  {author} {\bibinfo {author} {\bibfnamefont {T.~W.}\ \bibnamefont
			{Schmidt}},\ }\bibfield  {title} {\enquote {\bibinfo {title} {{A Marcus-Hush
					perspective on adiabatic singlet fission}},}\ }\href {\doibase
		10.1063/1.5108669} {\bibfield  {journal} {\bibinfo  {journal} {Journal of
				Chemical Physics}\ }\textbf {\bibinfo {volume} {151}} (\bibinfo {year}
		{2019}),\ 10.1063/1.5108669}\BibitemShut {NoStop}%
	\bibitem [{\citenamefont {Becke}(1993)}]{becke93a}%
	\BibitemOpen
	\bibfield  {author} {\bibinfo {author} {\bibfnamefont {A.~D.}\ \bibnamefont
			{Becke}},\ }\bibfield  {title} {\enquote {\bibinfo {title}
			{Density‐functional thermochemistry. iii. the role of exact exchange},}\
	}\href {\doibase 10.1063/1.464913} {\bibfield  {journal} {\bibinfo  {journal}
			{The Journal of Chemical Physics}\ }\textbf {\bibinfo {volume} {98}},\
		\bibinfo {pages} {5648--5652} (\bibinfo {year} {1993})},\ \Eprint
	{http://arxiv.org/abs/https://doi.org/10.1063/1.464913}
	{https://doi.org/10.1063/1.464913} \BibitemShut {NoStop}%
	\bibitem [{\citenamefont {Petersson}\ \emph {et~al.}(1988)\citenamefont
		{Petersson}, \citenamefont {Bennett}, \citenamefont {Tensfeldt},
		\citenamefont {Al‐Laham}, \citenamefont {Shirley},\ and\ \citenamefont
		{Mantzaris}}]{petersson88}%
	\BibitemOpen
	\bibfield  {author} {\bibinfo {author} {\bibfnamefont {G.~A.}\ \bibnamefont
			{Petersson}}, \bibinfo {author} {\bibfnamefont {A.}~\bibnamefont {Bennett}},
		\bibinfo {author} {\bibfnamefont {T.~G.}\ \bibnamefont {Tensfeldt}}, \bibinfo
		{author} {\bibfnamefont {M.~A.}\ \bibnamefont {Al‐Laham}}, \bibinfo
		{author} {\bibfnamefont {W.~A.}\ \bibnamefont {Shirley}}, \ and\ \bibinfo
		{author} {\bibfnamefont {J.}~\bibnamefont {Mantzaris}},\ }\bibfield  {title}
	{\enquote {\bibinfo {title} {A complete basis set model chemistry. i. the
				total energies of closed‐shell atoms and hydrides of the first‐row
				elements},}\ }\href {\doibase 10.1063/1.455064} {\bibfield  {journal}
		{\bibinfo  {journal} {The Journal of Chemical Physics}\ }\textbf {\bibinfo
			{volume} {89}},\ \bibinfo {pages} {2193--2218} (\bibinfo {year} {1988})},\
	\Eprint {http://arxiv.org/abs/https://doi.org/10.1063/1.455064}
	{https://doi.org/10.1063/1.455064} \BibitemShut {NoStop}%
	\bibitem [{\citenamefont {Petersson}\ and\ \citenamefont
		{Al‐Laham}(1991)}]{petersson91}%
	\BibitemOpen
	\bibfield  {author} {\bibinfo {author} {\bibfnamefont {G.~A.}\ \bibnamefont
			{Petersson}}\ and\ \bibinfo {author} {\bibfnamefont {M.~A.}\ \bibnamefont
			{Al‐Laham}},\ }\bibfield  {title} {\enquote {\bibinfo {title} {A complete
				basis set model chemistry. ii. open‐shell systems and the total energies of
				the first‐row atoms},}\ }\href {\doibase 10.1063/1.460447} {\bibfield
		{journal} {\bibinfo  {journal} {The Journal of Chemical Physics}\ }\textbf
		{\bibinfo {volume} {94}},\ \bibinfo {pages} {6081--6090} (\bibinfo {year}
		{1991})},\ \Eprint {http://arxiv.org/abs/https://doi.org/10.1063/1.460447}
	{https://doi.org/10.1063/1.460447} \BibitemShut {NoStop}%
	\bibitem [{\citenamefont {Frisch}\ \emph {et~al.}(2016)\citenamefont {Frisch},
		\citenamefont {Trucks}, \citenamefont {Schlegel}, \citenamefont {Scuseria},
		\citenamefont {Robb},\ and\ \citenamefont
		{Cheeseman}}]{Frisch2016GaussianC.01}%
	\BibitemOpen
	\bibfield  {author} {\bibinfo {author} {\bibfnamefont {M.~J.}\ \bibnamefont
			{Frisch}}, \bibinfo {author} {\bibfnamefont {G.~W.}\ \bibnamefont {Trucks}},
		\bibinfo {author} {\bibfnamefont {H.~B.}\ \bibnamefont {Schlegel}}, \bibinfo
		{author} {\bibfnamefont {G.~E.}\ \bibnamefont {Scuseria}}, \bibinfo {author}
		{\bibfnamefont {M.~A.}\ \bibnamefont {Robb}}, \ and\ \bibinfo {author}
		{\bibfnamefont {J.~R.}\ \bibnamefont {Cheeseman}},\ }\href@noop {} {\enquote
		{\bibinfo {title} {{Gaussian 16, Revision C.01}},}\ } (\bibinfo {year}
	{2016})\BibitemShut {NoStop}%
	\bibitem [{\citenamefont {K{\"{o}}hler}\ and\ \citenamefont
		{B{\"{a}}ssler}(2015)}]{Kohler2015ElectronicSemiconductors}%
	\BibitemOpen
	\bibfield  {author} {\bibinfo {author} {\bibfnamefont {A.}~\bibnamefont
			{K{\"{o}}hler}}\ and\ \bibinfo {author} {\bibfnamefont {H.}~\bibnamefont
			{B{\"{a}}ssler}},\ }\href {\doibase 10.1002/9783527685172} {\emph {\bibinfo
			{title} {{Electronic Processes in Organic Semiconductors}}}}\ (\bibinfo
	{publisher} {Wiley-VCH Verlag GmbH {\&} Co. KGaA},\ \bibinfo {address}
	{Weinheim, Germany},\ \bibinfo {year} {2015})\BibitemShut {NoStop}%
	\bibitem [{\citenamefont {Pun}\ \emph {et~al.}(2018)\citenamefont {Pun},
		\citenamefont {Gallaher}, \citenamefont {Frazer}, \citenamefont {Prasad},
		\citenamefont {Dover}, \citenamefont {MacQueen},\ and\ \citenamefont
		{Schmidt}}]{pun2018}%
	\BibitemOpen
	\bibfield  {author} {\bibinfo {author} {\bibfnamefont {J.~K.~H.}\
			\bibnamefont {Pun}}, \bibinfo {author} {\bibfnamefont {J.~K.}\ \bibnamefont
			{Gallaher}}, \bibinfo {author} {\bibfnamefont {L.}~\bibnamefont {Frazer}},
		\bibinfo {author} {\bibfnamefont {S.~K.~K.}\ \bibnamefont {Prasad}}, \bibinfo
		{author} {\bibfnamefont {C.~B.}\ \bibnamefont {Dover}}, \bibinfo {author}
		{\bibfnamefont {R.~W.}\ \bibnamefont {MacQueen}}, \ and\ \bibinfo {author}
		{\bibfnamefont {T.~W.}\ \bibnamefont {Schmidt}},\ }\bibfield  {title}
	{\enquote {\bibinfo {title} {{TIPS-anthracene: a singlet fission or triplet
					fusion material?}}}\ }\href {\doibase 10.1117/1.JPE.8.022006} {\bibfield
		{journal} {\bibinfo  {journal} {Journal of Photonics for Energy}\ }\textbf
		{\bibinfo {volume} {8}},\ \bibinfo {pages} {022006} (\bibinfo {year}
		{2018})}\BibitemShut {NoStop}%
	\bibitem [{\citenamefont {Jensen}(2007)}]{Jensen2007IntroductionChemistry}%
	\BibitemOpen
	\bibfield  {author} {\bibinfo {author} {\bibfnamefont {F.}~\bibnamefont
			{Jensen}},\ }\href@noop {} {\emph {\bibinfo {title} {{Introduction to
					Computational Chemistry}}}},\ \bibinfo {edition} {2nd}\ ed.\ (\bibinfo
	{publisher} {John Wiley {\&} Sons Ltd},\ \bibinfo {address} {West Sussex,
		England},\ \bibinfo {year} {2007})\BibitemShut {NoStop}%
	\bibitem [{\citenamefont {Imamura}\ and\ \citenamefont
		{Hoffmann}(1968)}]{Imamura1968TheCompounds}%
	\BibitemOpen
	\bibfield  {author} {\bibinfo {author} {\bibfnamefont {A.}~\bibnamefont
			{Imamura}}\ and\ \bibinfo {author} {\bibfnamefont {R.}~\bibnamefont
			{Hoffmann}},\ }\bibfield  {title} {\enquote {\bibinfo {title} {{The
					Electronic Structure and Torsional Potentials in Ground and Excited States of
					Biphenyl, Fulvalene, and Related Compounds}},}\ }\href {\doibase
		10.1021/ja01022a008} {\bibfield  {journal} {\bibinfo  {journal} {Journal of
				the American Chemical Society}\ }\textbf {\bibinfo {volume} {90}},\ \bibinfo
		{pages} {5379--5385} (\bibinfo {year} {1968})}\BibitemShut {NoStop}%
	\bibitem [{\citenamefont {Bastiansen}\ and\ \citenamefont
		{Samdel}(1985)}]{Bastiansen1985StructureDerivatives}%
	\BibitemOpen
	\bibfield  {author} {\bibinfo {author} {\bibfnamefont {O.}~\bibnamefont
			{Bastiansen}}\ and\ \bibinfo {author} {\bibfnamefont {S.}~\bibnamefont
			{Samdel}},\ }\bibfield  {title} {\enquote {\bibinfo {title} {{Structure and
					barrier of internal rotation of biphenyl derivatives in the gaseous state.
					Part 4. Barrier of internal rotation in biphenyl, perdeuterated biphenyl and
					seven non-ortho-substituted halogen derivatives}},}\ }\href
	{http://ac.els-cdn.com/0022286085850444/1-s2.0-0022286085850444-main.pdf?_tid=edb5f8e4-c7bb-11e5-b293-00000aacb361&acdnat=1454204746_a121a2e9caee699442a46f8cfb7695ba}
	{\bibfield  {journal} {\bibinfo  {journal} {Journal of Molecular Structure}\
		}\textbf {\bibinfo {volume} {128}},\ \bibinfo {pages} {115--125} (\bibinfo
		{year} {1985})}\BibitemShut {NoStop}%
	\bibitem [{\citenamefont {Benk}\ and\ \citenamefont
		{Sixl}(1981)}]{Benk1981TheoryStructures}%
	\BibitemOpen
	\bibfield  {author} {\bibinfo {author} {\bibfnamefont {H.}~\bibnamefont
			{Benk}}\ and\ \bibinfo {author} {\bibfnamefont {H.}~\bibnamefont {Sixl}},\
	}\bibfield  {title} {\enquote {\bibinfo {title} {{Theory of two coupled
					triplet states application to bicarbene structures}},}\ }\href {\doibase
		10.1080/00268978100100631} {\bibfield  {journal} {\bibinfo  {journal}
			{Molecular Physics}\ }\textbf {\bibinfo {volume} {42}},\ \bibinfo {pages}
		{779--801} (\bibinfo {year} {1981})}\BibitemShut {NoStop}%
	\bibitem [{\citenamefont {Schmidt}\ \emph {et~al.}(1993)\citenamefont
		{Schmidt}, \citenamefont {Baldridge}, \citenamefont {Boatz}, \citenamefont
		{Elbert}, \citenamefont {Gordon}, \citenamefont {Jensen}, \citenamefont
		{Koseki}, \citenamefont {Matsunaga}, \citenamefont {Nguyen}, \citenamefont
		{Su}, \citenamefont {Windus}, \citenamefont {Dupuis},\ and\ \citenamefont
		{Montgomery}}]{Schmidt1993GeneralSystem}%
	\BibitemOpen
	\bibfield  {author} {\bibinfo {author} {\bibfnamefont {M.~W.}\ \bibnamefont
			{Schmidt}}, \bibinfo {author} {\bibfnamefont {K.~K.}\ \bibnamefont
			{Baldridge}}, \bibinfo {author} {\bibfnamefont {J.~A.}\ \bibnamefont
			{Boatz}}, \bibinfo {author} {\bibfnamefont {S.~T.}\ \bibnamefont {Elbert}},
		\bibinfo {author} {\bibfnamefont {M.~S.}\ \bibnamefont {Gordon}}, \bibinfo
		{author} {\bibfnamefont {J.~H.}\ \bibnamefont {Jensen}}, \bibinfo {author}
		{\bibfnamefont {S.}~\bibnamefont {Koseki}}, \bibinfo {author} {\bibfnamefont
			{N.}~\bibnamefont {Matsunaga}}, \bibinfo {author} {\bibfnamefont {K.~A.}\
			\bibnamefont {Nguyen}}, \bibinfo {author} {\bibfnamefont {S.}~\bibnamefont
			{Su}}, \bibinfo {author} {\bibfnamefont {T.~L.}\ \bibnamefont {Windus}},
		\bibinfo {author} {\bibfnamefont {M.}~\bibnamefont {Dupuis}}, \ and\ \bibinfo
		{author} {\bibfnamefont {J.~A.}\ \bibnamefont {Montgomery}},\ }\bibfield
	{title} {\enquote {\bibinfo {title} {{General atomic and molecular electronic
					structure system}},}\ }\href {\doibase 10.1002/jcc.540141112} {\bibfield
		{journal} {\bibinfo  {journal} {Journal of Computational Chemistry}\ }\textbf
		{\bibinfo {volume} {14}},\ \bibinfo {pages} {1347--1363} (\bibinfo {year}
		{1993})}\BibitemShut {NoStop}%
	\bibitem [{\citenamefont {Rugg}\ \emph {et~al.}(2022)\citenamefont {Rugg},
		\citenamefont {Smyser}, \citenamefont {Fluegel}, \citenamefont {Chang},
		\citenamefont {Thorley}, \citenamefont {Parkin}, \citenamefont {Anthony},
		\citenamefont {Eaves},\ and\ \citenamefont {Johnson}}]{Rugg2022a}%
	\BibitemOpen
	\bibfield  {author} {\bibinfo {author} {\bibfnamefont {B.~K.}\ \bibnamefont
			{Rugg}}, \bibinfo {author} {\bibfnamefont {K.~E.}\ \bibnamefont {Smyser}},
		\bibinfo {author} {\bibfnamefont {B.}~\bibnamefont {Fluegel}}, \bibinfo
		{author} {\bibfnamefont {C.~H.}\ \bibnamefont {Chang}}, \bibinfo {author}
		{\bibfnamefont {K.~J.}\ \bibnamefont {Thorley}}, \bibinfo {author}
		{\bibfnamefont {S.}~\bibnamefont {Parkin}}, \bibinfo {author} {\bibfnamefont
			{J.~E.}\ \bibnamefont {Anthony}}, \bibinfo {author} {\bibfnamefont {J.~D.}\
			\bibnamefont {Eaves}}, \ and\ \bibinfo {author} {\bibfnamefont {J.~C.}\
			\bibnamefont {Johnson}},\ }\bibfield  {title} {\enquote {\bibinfo {title}
			{Triplet-pair spin signatures from macroscopically aligned heteroacenes in an
				oriented single crystal},}\ }\href {\doibase 10.1073/pnas.2201879119}
	{\bibfield  {journal} {\bibinfo  {journal} {Proc. Natl. Acad. Sci.}\ }\textbf
		{\bibinfo {volume} {119}},\ \bibinfo {pages} {e2201879119} (\bibinfo {year}
		{2022})}\BibitemShut {NoStop}%
	\bibitem [{\citenamefont {Hasobe}\ \emph {et~al.}(2022)\citenamefont {Hasobe},
		\citenamefont {Nakamura}, \citenamefont {Tkachenko},\ and\ \citenamefont
		{Kobori}}]{Hasobe2022}%
	\BibitemOpen
	\bibfield  {author} {\bibinfo {author} {\bibfnamefont {T.}~\bibnamefont
			{Hasobe}}, \bibinfo {author} {\bibfnamefont {S.}~\bibnamefont {Nakamura}},
		\bibinfo {author} {\bibfnamefont {N.~V.}\ \bibnamefont {Tkachenko}}, \ and\
		\bibinfo {author} {\bibfnamefont {Y.}~\bibnamefont {Kobori}},\ }\bibfield
	{title} {\enquote {\bibinfo {title} {Molecular {{Design Strategy}} for
				{{High-Yield}} and {{Long-Lived Individual Doubled Triplet Excitons}} through
				{{Intramolecular Singlet Fission}}},}\ }\href {\doibase
		10.1021/acsenergylett.1c02300} {\bibfield  {journal} {\bibinfo  {journal}
			{ACS Energy Lett.}\ }\textbf {\bibinfo {volume} {7}},\ \bibinfo {pages}
		{390--400} (\bibinfo {year} {2022})}\BibitemShut {NoStop}%
	\bibitem [{\citenamefont {Tiana}, \citenamefont {Sutto},\ and\ \citenamefont
		{Broglia}(2007)}]{Tiana2007UseChains}%
	\BibitemOpen
	\bibfield  {author} {\bibinfo {author} {\bibfnamefont {G.}~\bibnamefont
			{Tiana}}, \bibinfo {author} {\bibfnamefont {L.}~\bibnamefont {Sutto}}, \ and\
		\bibinfo {author} {\bibfnamefont {R.~A.}\ \bibnamefont {Broglia}},\
	}\bibfield  {title} {\enquote {\bibinfo {title} {{Use of the Metropolis
					algorithm to simulate the dynamics of protein chains}},}\ }\href {\doibase
		10.1016/j.physa.2007.02.044} {\bibfield  {journal} {\bibinfo  {journal}
			{Physica A: Statistical Mechanics and its Applications}\ }\textbf {\bibinfo
			{volume} {380}},\ \bibinfo {pages} {241--249} (\bibinfo {year}
		{2007})}\BibitemShut {NoStop}%
	\bibitem [{\citenamefont {Sakai}\ \emph {et~al.}(2018)\citenamefont {Sakai},
		\citenamefont {Inaya}, \citenamefont {Nagashima}, \citenamefont {Nakamura},
		\citenamefont {Kobori}, \citenamefont {Tkachenko},\ and\ \citenamefont
		{Hasobe}}]{Sakai2018MultiexcitonPairs}%
	\BibitemOpen
	\bibfield  {author} {\bibinfo {author} {\bibfnamefont {H.}~\bibnamefont
			{Sakai}}, \bibinfo {author} {\bibfnamefont {R.}~\bibnamefont {Inaya}},
		\bibinfo {author} {\bibfnamefont {H.}~\bibnamefont {Nagashima}}, \bibinfo
		{author} {\bibfnamefont {S.}~\bibnamefont {Nakamura}}, \bibinfo {author}
		{\bibfnamefont {Y.}~\bibnamefont {Kobori}}, \bibinfo {author} {\bibfnamefont
			{N.~V.}\ \bibnamefont {Tkachenko}}, \ and\ \bibinfo {author} {\bibfnamefont
			{T.}~\bibnamefont {Hasobe}},\ }\bibfield  {title} {\enquote {\bibinfo {title}
			{{Multiexciton Dynamics Depending on Intramolecular Orientations in Pentacene
					Dimers: Recombination and Dissociation of Correlated Triplet Pairs}},}\
	}\href {\doibase 10.1021/acs.jpclett.8b01184} {\bibfield  {journal} {\bibinfo
			{journal} {Journal of Physical Chemistry Letters}\ }\textbf {\bibinfo
			{volume} {9}},\ \bibinfo {pages} {3354--3360} (\bibinfo {year}
		{2018})}\BibitemShut {NoStop}%
	\bibitem [{\citenamefont {Pun}\ \emph {et~al.}(2019)\citenamefont {Pun},
		\citenamefont {Asadpoordarvish}, \citenamefont {Kumarasamy}, \citenamefont
		{Tayebjee}, \citenamefont {Niesner}, \citenamefont {McCamey}, \citenamefont
		{Sanders}, \citenamefont {Campos},\ and\ \citenamefont
		{Sfeir}}]{Pun2019Ultra-fastDesign}%
	\BibitemOpen
	\bibfield  {author} {\bibinfo {author} {\bibfnamefont {A.~B.}\ \bibnamefont
			{Pun}}, \bibinfo {author} {\bibfnamefont {A.}~\bibnamefont
			{Asadpoordarvish}}, \bibinfo {author} {\bibfnamefont {E.}~\bibnamefont
			{Kumarasamy}}, \bibinfo {author} {\bibfnamefont {M.~J.}\ \bibnamefont
			{Tayebjee}}, \bibinfo {author} {\bibfnamefont {D.}~\bibnamefont {Niesner}},
		\bibinfo {author} {\bibfnamefont {D.~R.}\ \bibnamefont {McCamey}}, \bibinfo
		{author} {\bibfnamefont {S.~N.}\ \bibnamefont {Sanders}}, \bibinfo {author}
		{\bibfnamefont {L.~M.}\ \bibnamefont {Campos}}, \ and\ \bibinfo {author}
		{\bibfnamefont {M.~Y.}\ \bibnamefont {Sfeir}},\ }\bibfield  {title} {\enquote
		{\bibinfo {title} {{Ultra-fast intramolecular singlet fission to persistent
					multiexcitons by molecular design}},}\ }\href {\doibase
		10.1038/s41557-019-0297-7} {\bibfield  {journal} {\bibinfo  {journal} {Nature
				Chemistry}\ }\textbf {\bibinfo {volume} {11}},\ \bibinfo {pages} {821--828}
		(\bibinfo {year} {2019})}\BibitemShut {NoStop}%
	\bibitem [{\citenamefont {Concistr{\`{e}}}\ \emph {et~al.}(2014)\citenamefont
		{Concistr{\`{e}}}, \citenamefont {Carignani}, \citenamefont {Borsacchi},
		\citenamefont {Johannessen}, \citenamefont {Mennucci}, \citenamefont {Yang},
		\citenamefont {Geppi},\ and\ \citenamefont
		{Levitt}}]{Concistre2014FreezingNMR}%
	\BibitemOpen
	\bibfield  {author} {\bibinfo {author} {\bibfnamefont {M.}~\bibnamefont
			{Concistr{\`{e}}}}, \bibinfo {author} {\bibfnamefont {E.}~\bibnamefont
			{Carignani}}, \bibinfo {author} {\bibfnamefont {S.}~\bibnamefont
			{Borsacchi}}, \bibinfo {author} {\bibfnamefont {O.~G.}\ \bibnamefont
			{Johannessen}}, \bibinfo {author} {\bibfnamefont {B.}~\bibnamefont
			{Mennucci}}, \bibinfo {author} {\bibfnamefont {Y.}~\bibnamefont {Yang}},
		\bibinfo {author} {\bibfnamefont {M.}~\bibnamefont {Geppi}}, \ and\ \bibinfo
		{author} {\bibfnamefont {M.~H.}\ \bibnamefont {Levitt}},\ }\bibfield  {title}
	{\enquote {\bibinfo {title} {{Freezing of molecular motions probed by
					cryogenic magic angle spinning NMR}},}\ }\href {\doibase 10.1021/jz4026276}
	{\bibfield  {journal} {\bibinfo  {journal} {Journal of Physical Chemistry
				Letters}\ }\textbf {\bibinfo {volume} {5}},\ \bibinfo {pages} {512--516}
		(\bibinfo {year} {2014})}\BibitemShut {NoStop}%
	\bibitem [{\citenamefont {Illig}\ \emph {et~al.}(2016)\citenamefont {Illig},
		\citenamefont {Eggeman}, \citenamefont {Troisi}, \citenamefont {Jiang},
		\citenamefont {Warwick}, \citenamefont {Nikolka}, \citenamefont {Schweicher},
		\citenamefont {Yeates}, \citenamefont {Henri~Geerts}, \citenamefont
		{Anthony},\ and\ \citenamefont {Sirringhaus}}]{Illig2016ReducingMotions}%
	\BibitemOpen
	\bibfield  {author} {\bibinfo {author} {\bibfnamefont {S.}~\bibnamefont
			{Illig}}, \bibinfo {author} {\bibfnamefont {A.~S.}\ \bibnamefont {Eggeman}},
		\bibinfo {author} {\bibfnamefont {A.}~\bibnamefont {Troisi}}, \bibinfo
		{author} {\bibfnamefont {L.}~\bibnamefont {Jiang}}, \bibinfo {author}
		{\bibfnamefont {C.}~\bibnamefont {Warwick}}, \bibinfo {author} {\bibfnamefont
			{M.}~\bibnamefont {Nikolka}}, \bibinfo {author} {\bibfnamefont
			{G.}~\bibnamefont {Schweicher}}, \bibinfo {author} {\bibfnamefont {S.~G.}\
			\bibnamefont {Yeates}}, \bibinfo {author} {\bibfnamefont {Y.}~\bibnamefont
			{Henri~Geerts}}, \bibinfo {author} {\bibfnamefont {J.~E.}\ \bibnamefont
			{Anthony}}, \ and\ \bibinfo {author} {\bibfnamefont {H.}~\bibnamefont
			{Sirringhaus}},\ }\bibfield  {title} {\enquote {\bibinfo {title} {{Reducing
					dynamic disorder in small-molecule organic semiconductors by suppressing
					large-Amplitude thermal motions}},}\ }\href {\doibase 10.1038/ncomms10736}
	{\bibfield  {journal} {\bibinfo  {journal} {Nature Communications}\ }\textbf
		{\bibinfo {volume} {7}},\ \bibinfo {pages} {1--10} (\bibinfo {year}
		{2016})}\BibitemShut {NoStop}%
	\bibitem [{\citenamefont {Kwon}\ \emph {et~al.}(2015)\citenamefont {Kwon},
		\citenamefont {Yu}, \citenamefont {Coburn}, \citenamefont {Phillips},
		\citenamefont {Chung}, \citenamefont {Shanker}, \citenamefont {Jung},
		\citenamefont {Kim}, \citenamefont {Pipe}, \citenamefont {Forrest},
		\citenamefont {Youk}, \citenamefont {Gierschner},\ and\ \citenamefont
		{Kim}}]{Kwon2015SuppressingMaterials}%
	\BibitemOpen
	\bibfield  {author} {\bibinfo {author} {\bibfnamefont {M.~S.}\ \bibnamefont
			{Kwon}}, \bibinfo {author} {\bibfnamefont {Y.}~\bibnamefont {Yu}}, \bibinfo
		{author} {\bibfnamefont {C.}~\bibnamefont {Coburn}}, \bibinfo {author}
		{\bibfnamefont {A.~W.}\ \bibnamefont {Phillips}}, \bibinfo {author}
		{\bibfnamefont {K.}~\bibnamefont {Chung}}, \bibinfo {author} {\bibfnamefont
			{A.}~\bibnamefont {Shanker}}, \bibinfo {author} {\bibfnamefont
			{J.}~\bibnamefont {Jung}}, \bibinfo {author} {\bibfnamefont {G.}~\bibnamefont
			{Kim}}, \bibinfo {author} {\bibfnamefont {K.}~\bibnamefont {Pipe}}, \bibinfo
		{author} {\bibfnamefont {S.~R.}\ \bibnamefont {Forrest}}, \bibinfo {author}
		{\bibfnamefont {J.~H.}\ \bibnamefont {Youk}}, \bibinfo {author}
		{\bibfnamefont {J.}~\bibnamefont {Gierschner}}, \ and\ \bibinfo {author}
		{\bibfnamefont {J.}~\bibnamefont {Kim}},\ }\bibfield  {title} {\enquote
		{\bibinfo {title} {{Suppressing molecular motions for enhanced roomerature
					phosphorescence of metal-free organic materials}},}\ }\href {\doibase
		10.1038/ncomms9947} {\bibfield  {journal} {\bibinfo  {journal} {Nature
				Communications}\ }\textbf {\bibinfo {volume} {6}} (\bibinfo {year} {2015}),\
		10.1038/ncomms9947}\BibitemShut {NoStop}%
	\bibitem [{\citenamefont {Nobukawa}\ \emph {et~al.}(2013)\citenamefont
		{Nobukawa}, \citenamefont {Urakawa}, \citenamefont {Shikata},\ and\
		\citenamefont {Inoue}}]{Nobukawa2013DynamicsTransition}%
	\BibitemOpen
	\bibfield  {author} {\bibinfo {author} {\bibfnamefont {S.}~\bibnamefont
			{Nobukawa}}, \bibinfo {author} {\bibfnamefont {O.}~\bibnamefont {Urakawa}},
		\bibinfo {author} {\bibfnamefont {T.}~\bibnamefont {Shikata}}, \ and\
		\bibinfo {author} {\bibfnamefont {T.}~\bibnamefont {Inoue}},\ }\bibfield
	{title} {\enquote {\bibinfo {title} {{Dynamics of a probe molecule dissolved
					in several polymer matrices with different side-chain structures:
					Determination of correlation length relevant to glass transition}},}\ }\href
	{\doibase 10.1021/ma302567j} {\bibfield  {journal} {\bibinfo  {journal}
			{Macromolecules}\ }\textbf {\bibinfo {volume} {46}},\ \bibinfo {pages}
		{2206--2215} (\bibinfo {year} {2013})}\BibitemShut {NoStop}%
	\bibitem [{\citenamefont {Johansson}, \citenamefont {Nation},\ and\
		\citenamefont {Nori}(2012)}]{Johansson2012QuTiP:Systems}%
	\BibitemOpen
	\bibfield  {author} {\bibinfo {author} {\bibfnamefont {J.}~\bibnamefont
			{Johansson}}, \bibinfo {author} {\bibfnamefont {P.}~\bibnamefont {Nation}}, \
		and\ \bibinfo {author} {\bibfnamefont {F.}~\bibnamefont {Nori}},\ }\bibfield
	{title} {\enquote {\bibinfo {title} {{QuTiP: An open-source Python framework
					for the dynamics of open quantum systems}},}\ }\href {\doibase
		10.1016/j.cpc.2012.02.021} {\bibfield  {journal} {\bibinfo  {journal}
			{Computer Physics Communications}\ }\textbf {\bibinfo {volume} {183}},\
		\bibinfo {pages} {1760--1772} (\bibinfo {year} {2012})}\BibitemShut {NoStop}%
	\bibitem [{\citenamefont {Bayliss}\ \emph {et~al.}(2016)\citenamefont
		{Bayliss}, \citenamefont {Weiss}, \citenamefont {Rao}, \citenamefont
		{Friend}, \citenamefont {Chepelianskii},\ and\ \citenamefont
		{Greenham}}]{Bayliss2016SpinFission}%
	\BibitemOpen
	\bibfield  {author} {\bibinfo {author} {\bibfnamefont {S.~L.}\ \bibnamefont
			{Bayliss}}, \bibinfo {author} {\bibfnamefont {L.~R.}\ \bibnamefont {Weiss}},
		\bibinfo {author} {\bibfnamefont {A.}~\bibnamefont {Rao}}, \bibinfo {author}
		{\bibfnamefont {R.~H.}\ \bibnamefont {Friend}}, \bibinfo {author}
		{\bibfnamefont {A.~D.}\ \bibnamefont {Chepelianskii}}, \ and\ \bibinfo
		{author} {\bibfnamefont {N.~C.}\ \bibnamefont {Greenham}},\ }\bibfield
	{title} {\enquote {\bibinfo {title} {{Spin signatures of exchange-coupled
					triplet pairs formed by singlet fission}},}\ }\href {\doibase
		10.1103/PhysRevB.94.045204} {\bibfield  {journal} {\bibinfo  {journal}
			{Physical Review B}\ }\textbf {\bibinfo {volume} {94}},\ \bibinfo {pages}
		{1--7} (\bibinfo {year} {2016})}\BibitemShut {NoStop}%
	\bibitem [{\citenamefont {MacDonald}\ \emph {et~al.}(2023)\citenamefont
		{MacDonald}, \citenamefont {Tayebjee}, \citenamefont {Collins}, \citenamefont
		{Kumarasamy}, \citenamefont {Sanders}, \citenamefont {Sfeir}, \citenamefont
		{Campos},\ and\ \citenamefont
		{McCamey}}]{macdonald_tayebjee_collins_kumarasamy_sanders_sfeir_campos_mccamey_2023}%
	\BibitemOpen
	\bibfield  {author} {\bibinfo {author} {\bibfnamefont {T.}~\bibnamefont
			{MacDonald}}, \bibinfo {author} {\bibfnamefont {M.}~\bibnamefont {Tayebjee}},
		\bibinfo {author} {\bibfnamefont {M.}~\bibnamefont {Collins}}, \bibinfo
		{author} {\bibfnamefont {E.}~\bibnamefont {Kumarasamy}}, \bibinfo {author}
		{\bibfnamefont {S.}~\bibnamefont {Sanders}}, \bibinfo {author} {\bibfnamefont
			{M.}~\bibnamefont {Sfeir}}, \bibinfo {author} {\bibfnamefont
			{L.}~\bibnamefont {Campos}}, \ and\ \bibinfo {author} {\bibfnamefont
			{D.}~\bibnamefont {McCamey}},\ }\bibfield  {title} {\enquote {\bibinfo
			{title} {Anisotropic multiexciton quintet and triplet dynamics in singlet
				fission via peanut},}\ }\href {\doibase 10.26434/chemrxiv-2023-n60mv}
	{\bibfield  {journal} {\bibinfo  {journal} {ChemRxiv}\ } (\bibinfo {year}
		{2023}),\ 10.26434/chemrxiv-2023-n60mv}\BibitemShut {NoStop}%
	\bibitem [{\citenamefont {Nakamura}\ \emph {et~al.}(2021)\citenamefont
		{Nakamura}, \citenamefont {Sakai}, \citenamefont {Nagashima}, \citenamefont
		{Fuki}, \citenamefont {Onishi}, \citenamefont {Khan}, \citenamefont {Kobori},
		\citenamefont {Tkachenko},\ and\ \citenamefont
		{Hasobe}}]{Nakamura2021SynergeticFission}%
	\BibitemOpen
	\bibfield  {author} {\bibinfo {author} {\bibfnamefont {S.}~\bibnamefont
			{Nakamura}}, \bibinfo {author} {\bibfnamefont {H.}~\bibnamefont {Sakai}},
		\bibinfo {author} {\bibfnamefont {H.}~\bibnamefont {Nagashima}}, \bibinfo
		{author} {\bibfnamefont {M.}~\bibnamefont {Fuki}}, \bibinfo {author}
		{\bibfnamefont {K.}~\bibnamefont {Onishi}}, \bibinfo {author} {\bibfnamefont
			{R.}~\bibnamefont {Khan}}, \bibinfo {author} {\bibfnamefont {Y.}~\bibnamefont
			{Kobori}}, \bibinfo {author} {\bibfnamefont {N.~V.}\ \bibnamefont
			{Tkachenko}}, \ and\ \bibinfo {author} {\bibfnamefont {T.}~\bibnamefont
			{Hasobe}},\ }\bibfield  {title} {\enquote {\bibinfo {title} {{Synergetic Role
					of Conformational Flexibility and Electronic Coupling for Quantitative
					Intramolecular Singlet Fission}},}\ }\href {\doibase
		10.1021/acs.jpcc.1c04734} {\bibfield  {journal} {\bibinfo  {journal} {Journal
				of Physical Chemistry C}\ }\textbf {\bibinfo {volume} {125}},\ \bibinfo
		{pages} {18287--18296} (\bibinfo {year} {2021})}\BibitemShut {NoStop}%
	\bibitem [{\citenamefont {Chalyavi}\ \emph {et~al.}(2012)\citenamefont
		{Chalyavi}, \citenamefont {Troy}, \citenamefont {Bacskay}, \citenamefont
		{Nauta}, \citenamefont {Kable}, \citenamefont {Reid},\ and\ \citenamefont
		{Schmidt}}]{Chalyavi2012ExcitationRadicals}%
	\BibitemOpen
	\bibfield  {author} {\bibinfo {author} {\bibfnamefont {N.}~\bibnamefont
			{Chalyavi}}, \bibinfo {author} {\bibfnamefont {T.~P.}\ \bibnamefont {Troy}},
		\bibinfo {author} {\bibfnamefont {G.~B.}\ \bibnamefont {Bacskay}}, \bibinfo
		{author} {\bibfnamefont {K.}~\bibnamefont {Nauta}}, \bibinfo {author}
		{\bibfnamefont {S.~H.}\ \bibnamefont {Kable}}, \bibinfo {author}
		{\bibfnamefont {S.~A.}\ \bibnamefont {Reid}}, \ and\ \bibinfo {author}
		{\bibfnamefont {T.~W.}\ \bibnamefont {Schmidt}},\ }\bibfield  {title}
	{\enquote {\bibinfo {title} {{Excitation spectra of the jet-cooled
					4-phenylbenzyl and (methylphenyl)benzyl radicals}},}\ }\href {\doibase
		10.1021/jp309003u} {\bibfield  {journal} {\bibinfo  {journal} {Journal of
				Physical Chemistry A}\ }\textbf {\bibinfo {volume} {116}},\ \bibinfo {pages}
		{10780--10785} (\bibinfo {year} {2012})}\BibitemShut {NoStop}%
	\bibitem [{\citenamefont {Traiphol}\ \emph {et~al.}(2007)\citenamefont
		{Traiphol}, \citenamefont {Srikhirin}, \citenamefont {Kerdcharoen},
		\citenamefont {Osotchan}, \citenamefont {Scharnagl},\ and\ \citenamefont
		{Willumeit}}]{Traiphol2007InfluencesPolymer}%
	\BibitemOpen
	\bibfield  {author} {\bibinfo {author} {\bibfnamefont {R.}~\bibnamefont
			{Traiphol}}, \bibinfo {author} {\bibfnamefont {T.}~\bibnamefont {Srikhirin}},
		\bibinfo {author} {\bibfnamefont {T.}~\bibnamefont {Kerdcharoen}}, \bibinfo
		{author} {\bibfnamefont {T.}~\bibnamefont {Osotchan}}, \bibinfo {author}
		{\bibfnamefont {N.}~\bibnamefont {Scharnagl}}, \ and\ \bibinfo {author}
		{\bibfnamefont {R.}~\bibnamefont {Willumeit}},\ }\bibfield  {title} {\enquote
		{\bibinfo {title} {{Influences of local polymer-solvent
					{$\pi$}-{$\pi$}-interaction on dynamics of phenyl ring rotation and its role
					on photophysics of conjugated polymer}},}\ }\href {\doibase
		10.1016/j.eurpolymj.2006.11.022} {\bibfield  {journal} {\bibinfo  {journal}
			{European Polymer Journal}\ }\textbf {\bibinfo {volume} {43}},\ \bibinfo
		{pages} {478--487} (\bibinfo {year} {2007})}\BibitemShut {NoStop}%
	\bibitem [{\citenamefont {Banerjee}, \citenamefont {Yashonath},\ and\
		\citenamefont {Bagchi}(2016)}]{Banerjee2016CoupledSolutions}%
	\BibitemOpen
	\bibfield  {author} {\bibinfo {author} {\bibfnamefont {P.}~\bibnamefont
			{Banerjee}}, \bibinfo {author} {\bibfnamefont {S.}~\bibnamefont {Yashonath}},
		\ and\ \bibinfo {author} {\bibfnamefont {B.}~\bibnamefont {Bagchi}},\
	}\bibfield  {title} {\enquote {\bibinfo {title} {{Coupled jump rotational
					dynamics in aqueous nitrate solutions}},}\ }\href {\doibase
		10.1063/1.4971864} {\bibfield  {journal} {\bibinfo  {journal} {The Journal of
				Chemical Physics}\ }\textbf {\bibinfo {volume} {145}},\ \bibinfo {pages}
		{234502} (\bibinfo {year} {2016})}\BibitemShut {NoStop}%
	\bibitem [{\citenamefont {Mousseau}\ and\ \citenamefont
		{Barkema}(1998)}]{Mousseau1998TravelingTechnique}%
	\BibitemOpen
	\bibfield  {author} {\bibinfo {author} {\bibfnamefont {N.}~\bibnamefont
			{Mousseau}}\ and\ \bibinfo {author} {\bibfnamefont {G.~T.}\ \bibnamefont
			{Barkema}},\ }\bibfield  {title} {\enquote {\bibinfo {title} {{Traveling
					through potential energy landscapes of disordered materials: The
					activation-relaxation technique}},}\ }\href {\doibase
		10.1103/PhysRevE.57.2419} {\bibfield  {journal} {\bibinfo  {journal}
			{Physical Review E - Statistical Physics, Plasmas, Fluids, and Related
				Interdisciplinary Topics}\ }\textbf {\bibinfo {volume} {57}},\ \bibinfo
		{pages} {2419--2424} (\bibinfo {year} {1998})}\BibitemShut {NoStop}%
	\bibitem [{\citenamefont {Landau}\ and\ \citenamefont
		{Binder}(2005)}]{Landau2005APhysics}%
	\BibitemOpen
	\bibfield  {author} {\bibinfo {author} {\bibfnamefont {D.~P.}\ \bibnamefont
			{Landau}}\ and\ \bibinfo {author} {\bibfnamefont {K.}~\bibnamefont
			{Binder}},\ }\href {\doibase 10.1017/CBO9780511614460} {\emph {\bibinfo
			{title} {{A Guide to Monte Carlo Simulations in Statistical Physics}}}},\
	\bibinfo {edition} {2nd}\ ed.\ (\bibinfo  {publisher} {Cambridge University
		Press},\ \bibinfo {address} {Cambridge},\ \bibinfo {year} {2005})\BibitemShut
	{NoStop}%
	\bibitem [{\citenamefont {Berkelbach}, \citenamefont {Hybertsen},\ and\
		\citenamefont {Reichman}(2013)}]{Berkelbach2013MicroscopicFormulation}%
	\BibitemOpen
	\bibfield  {author} {\bibinfo {author} {\bibfnamefont {T.~C.}\ \bibnamefont
			{Berkelbach}}, \bibinfo {author} {\bibfnamefont {M.~S.}\ \bibnamefont
			{Hybertsen}}, \ and\ \bibinfo {author} {\bibfnamefont {D.~R.}\ \bibnamefont
			{Reichman}},\ }\bibfield  {title} {\enquote {\bibinfo {title} {{Microscopic
					theory of singlet exciton fission. I. General formulation}},}\ }\href
	{\doibase 10.1063/1.4794425} {\bibfield  {journal} {\bibinfo  {journal}
			{Journal of Chemical Physics}\ }\textbf {\bibinfo {volume} {138}} (\bibinfo
		{year} {2013}),\ 10.1063/1.4794425}\BibitemShut {NoStop}%
	\bibitem [{\citenamefont {Breuer}\ and\ \citenamefont
		{Petruccione}(2002)}]{Breuer2002TheSystems}%
	\BibitemOpen
	\bibfield  {author} {\bibinfo {author} {\bibfnamefont {H.-P.}\ \bibnamefont
			{Breuer}}\ and\ \bibinfo {author} {\bibfnamefont {F.}~\bibnamefont
			{Petruccione}},\ }\href@noop {} {\emph {\bibinfo {title} {{The Theory of Open
					Quantum Systems}}}}\ (\bibinfo  {publisher} {Oxford University Press},\
	\bibinfo {address} {Oxford},\ \bibinfo {year} {2002})\BibitemShut {NoStop}%
	\bibitem [{\citenamefont {Casanova}(2018)}]{Casanova2018TheoreticalFission}%
	\BibitemOpen
	\bibfield  {author} {\bibinfo {author} {\bibfnamefont {D.}~\bibnamefont
			{Casanova}},\ }\bibfield  {title} {\enquote {\bibinfo {title} {{Theoretical
					Modeling of Singlet Fission}},}\ }\href {\doibase
		10.1021/acs.chemrev.7b00601} {\bibfield  {journal} {\bibinfo  {journal}
			{Chemical Reviews}\ }\textbf {\bibinfo {volume} {118}},\ \bibinfo {pages}
		{7164--7207} (\bibinfo {year} {2018})}\BibitemShut {NoStop}%
	\bibitem [{\citenamefont {Metropolis}\ \emph {et~al.}(1953)\citenamefont
		{Metropolis}, \citenamefont {Rosenbluth}, \citenamefont {Rosenbluth},
		\citenamefont {Teller},\ and\ \citenamefont
		{Teller}}]{Metropolis1953EquationMachines}%
	\BibitemOpen
	\bibfield  {author} {\bibinfo {author} {\bibfnamefont {N.}~\bibnamefont
			{Metropolis}}, \bibinfo {author} {\bibfnamefont {A.~W.}\ \bibnamefont
			{Rosenbluth}}, \bibinfo {author} {\bibfnamefont {M.~N.}\ \bibnamefont
			{Rosenbluth}}, \bibinfo {author} {\bibfnamefont {A.~H.}\ \bibnamefont
			{Teller}}, \ and\ \bibinfo {author} {\bibfnamefont {E.}~\bibnamefont
			{Teller}},\ }\bibfield  {title} {\enquote {\bibinfo {title} {{Equation of
					State Calculations by Fast Computing Machines}},}\ }\href {\doibase
		10.1063/1.1699114} {\bibfield  {journal} {\bibinfo  {journal} {The Journal of
				Chemical Physics}\ }\textbf {\bibinfo {volume} {21}},\ \bibinfo {pages}
		{1087--1092} (\bibinfo {year} {1953})}\BibitemShut {NoStop}%
\end{thebibliography}
%

\clearpage
\newpage
\appendix
\onecolumngrid
\section{Determining Exchange}
\label{sec:SI-Exchange}
The GAMESS input files for the ROHF and CI calculations are shown in Code Excerpts~\ref{lst:orbs CI} and \ref{lst:CI input} respectively. The outputs of this calculation were the energies of the lowest 12 $S_Z=0$ electronic configurations across the molecular geometry space of the calculations an example of which is shown in \cref{tab:CI out}. In this typical output, the lowest three energy states can be identified as the ground singlet state $S_0S_0$, and the two singly excited triplet states, $\frac{1}{\sqrt{2}}(T_1S_0+S_0T_1)$ and $\frac{1}{\sqrt{2}}(T_1S_0-S_0T_1)$, with $\hat{S}=0,1,1$ respectively. The coupled triplet pairs that form the net singlet, triplet and quintet states, $^1(TT)_0, ^3(TT)_0$ and $^5(TT)_0$, were then identified as states $4,5$ and $6$. The value for $J$ for a particular geometry was then calculated as the energy difference of state 4 and 6 from GAMESS CI output. The sign of $J$ was determined by whether the $\hat{S}=2$ state was higher or lower (positive or negative $J$) in energy with respect to the $\hat{S}=0$ state.
\begin{table}[ht]
\centering
\begin{tabular}{ | m{2cm} | m{3cm}| m{3cm}| m{3cm} | m{3cm}|}
\hline
\textbf{STATE}   & \textbf{ENERGY ($E_h$)} & \textbf{SPIN}  & $\mathbf{S_z}$                                 \\
\hline
1 &  -2213.6207827296 & 0.00 & 0.00 \\
\hline
2 &  -2213.6081174362  & 1.00 & 0.00 \\
\hline
3 &  -2213.6081009926   & 1.00 &  0.00 \\
\hline 
4 &  -2213.5954441562   & 0.00 & 0.00 \\
\hline 
5 &  -2213.5954398750   & 1.00 & 0.00 \\
\hline 
6 &  -2213.5954313346   & 2.00 & 0.00 \\
$\vdots$ & $\vdots$  & $\vdots$ & $\vdots$  \\
\hline 
\end{tabular}
\caption{Example output from the CI calculation}
\label{tab:CI out}
\end{table}
\begin{lstlisting}[caption={GAMESS input file to determine the orbital occupancy for the quintet state},label={lst:orbs CI}]
#ROHF Calculation to optimize the orbitals for quintet states

 $CONTRL SCFTYP=ROHF RUNTYP=ENERGY
 MULT=5 ICHARG=0 MAXIT=70
 $END 
 $DET DISTCI=4 $END
 $TRANS MPTRAN=2 DIRTRF=.T. AOINTS=DIST $END
 $MCSCF FULLNR=.F. SOSCF=.T. FOCAS=.F. $END 
 $P2P P2P=.T. DLB=.T. $END
 $SCF DIIS=.T. SOSCF=.F. $END
 $SYSTEM TIMLIM=200 MWORDS=200 $END
 $BASIS GBASIS=N31 NGAUSS=6 NDFUNC=1 $END
 $GUESS GUESS=HUCKEL $END

 $DATA
 C58H30
C1
CARBON       6.0       -7.682556   -3.647399    0.969669
CARBON       6.0       -8.027450   -2.518207    0.758059
.            .          .           .           .
.            .          .           .           .
.            .          .           .           .
HYDROGEN     1.0        1.581654    1.973303   -1.961982
 $END
\end{lstlisting}

\begin{lstlisting}[caption={Configuration Interaction Calculation with a (4,4) active space (four electrons constrained to the four highest energy orbitals of the ROHF quintet).},label={lst:CI input}]
 $SYSTEM TIMLIM=120 MWORDS=500 MEMDDI=8275 $END
 $CONTRL SCFTYP=NONE RUNTYP=ENERGY
 MULT=1 ICHARG=0 NOSYM=1
 CITYP=ALDET
 $END 
! INTTYP=HONDO ICUT=11 ITOL=30
 $DET DISTCI=4 $END
 $TRANS MPTRAN=2 DIRTRF=.T. AOINTS=DIST $END
 $MCSCF FULLNR=.F. SOSCF=.T. FOCAS=.F. NORB=191 $END 
 $P2P P2P=.T. DLB=.T. $END
 $SCF DIIS=.T. SOSCF=.F. $END
 $GUESS GUESS=MOREAD NORB=250 $END
 $CIDET NCORE=187 NACT=4 NELS=4 NSTATE=12 $END

 $DATA  
 C58H30                                                                         
C1       0
CARBON      6.0     -8.1194440000      3.3501300000      1.5728110000
   N31     6
   D       1
     1              0.8000000000  1.00000000
.
.
.
 $END

--- OPEN SHELL ORBITALS --- GENERATED AT 12:29:02 12-SEP-2021    
 C58H30                                                                         
E(ROHF)=    -2213.5581888708, E(NUC)= 5863.4923252874,   17 ITERS
 $VEC   
 1  1-4.69803246E-09-8.95132119E-09-4.64815196E-09-4.38787828E-09 7.06539514E-10
 .
 .
 .
 $END
 POPULATION ANALYSIS
CARBON       6.53685  -0.53685   6.25662  -0.25662
.            .          .           .           .
.            .          .           .           .
.            .          .           .           .
HYDROGEN     1.0        1.581654    1.973303   -1.961982

 MOMENTS AT POINT    1 X,Y,Z= -0.000263  0.006091 -0.002655
 DIPOLE       0.064093 -0.101306  0.075525
\end{lstlisting}
\clearpage
\section{Metropolis Monte Carlo Simulations}
\label{sec:SI-MC}
\subsection{Selection of the Metropolis Monte Carlo Method}
The pentacene dimer \textbf{BP1} has two principle degrees of internal rotational freedom about the its single (1,4)-phenylene bridge. The single carbon-carbon bond linking each pentacene chromophore to the phenylene bridge is analogous to the well described internal degree of freedom of biphenyl\cite{Imamura1968TheCompounds}. Hence, we use the known behaviour of biphenyl to inform our model of \textbf{BP1} molecular dynamics.

Gas phase biphenyl dynamics are well described by quantised oscillations within and between the potential wells of the biphenyl potential energy surface \cite{Chalyavi2012ExcitationRadicals}. Because \textbf{BP1} singlet fission measurements are in general completed in solution or thin films~\cite{Tayebjee2017QuintetFission,Weiss2017StronglySemiconductor} these oscillations are likely to be rapidly damped by interactions with the solvent \cite{Traiphol2007InfluencesPolymer}. Addititionaly, the rotation of the molecule is strongly coupled to the phonon bath of the solution and will proceed in stochastic intra/inter-potential well jumps~\cite{Banerjee2016CoupledSolutions}. Considering the biphenyl energy surfaces and observations of dynamics in solution, we can construct some essential conditions. These conditions are the minimum that must be met to realistically sample from the internal rotational space of \textbf{BP1} over time.
\begin{enumerate}[label=(\roman*)]
    \item If the system starts in a non equilibrium conformation it will relax back to the energetic minima over some characteristic time.
    \item At equilibrium, occupation of energy states will be distributed according to the Boltzmann distribution.
    \item The system will stochastically `jump' from one orientation to the next with the transition rate dependent on the energy gap between the two orientations (size of phonon required).
    \item Large rotations will be inhibited by the solution and thus rare.
\end{enumerate}
Standard molecular dynamics techniques involve the integration over the equations of motion of the system~\cite{Mousseau1998TravelingTechnique}. For molecules containing hundreds of atoms, this approach is extremely resource intensive and limits the possible simulation time to $\sim 10^{-9}$ seconds \cite{Landau2005APhysics}. Further, the behaviour of molecules in solution is dominated by interactions with the environment ~\cite{Berkelbach2013MicroscopicFormulation}. This renders the problem more tractable to an open quantum systems approach where the exchange of phonons is explicitly dealt with~\cite{Breuer2002TheSystems,Casanova2018TheoreticalFission}. The open quantum systems approach requires a degree of complexity in its treatment of the phonon bath that is outside the scope of this paper. Without experimental testing or theoretical modelling of key parameters such as the solution density of states, the assumptions required also limit this approach.

In contrast, the Metropolis algorithm allows for the phonon bath to be treated implicitly~\cite{Landau2005APhysics}. The algorithm has been used widely to simulate stochastic trajectories and can couple the potential energy landscape to rotation probability, implicitly playing the role of a heat bath \cite{Landau2005APhysics}. It can be summarised by the following~\cite{Metropolis1953EquationMachines}:
\begin{enumerate}
    \item Choose an initial state $x_i$. 
    \item Wait an amount of time $t_i$ distributed around some characteristic time $\tau$. $\tau$ represents the average amount of time between jumps, the time taken to absorb or emit a phonon.
    \item Randomly choose a new state to step to $x_i' = x_i +\delta$. The distribution will be problem specific, and likely a free parameter to optimise.
    \item Calculate the energy change $\Delta U$ which results from this displacement. If $\Delta U < 0$ the move is accepted, $x_{i+1} = x_i'$. Return to step (2).
    \item If $\Delta U > 0$, choose a random number $\eta$ uniformly from $0 < \eta < 1$
    \item If $\eta < e^{-\Delta U/k_BT}$, accept the move. Here $T$ is the temperature and $k_B$ is the Boltzmann constant. Else if $\eta \geq e^{-\Delta U/k_BT}$, reject the move and $x_{i+1} = x_i$. Return to step (2).
\end{enumerate}
The properties of this algorithm have been extensively studied and it meets many of the requirements outlined above. Firstly, when a system is in equilibrium, the ensemble average of the orientations satisfies the Boltzmann distribution by construction~\cite{Metropolis1953EquationMachines}. This satisfies requirement (i) and (ii). The method also clearly meets requirement (iii) due to the energy dependencies of step (5) and (6). Work by Tiana et al. has demonstrated that under specific conditions the Metropolis algorithm can reproduce the dynamics of similar stochastic systems~\cite{Tiana2007UseChains}, supporting the viability of modelling \textbf{BP1} in this way. 
\subsection{Model Implementation}
\begin{figure}[ht]
    \centering
    \includegraphics[width = 0.7 \textwidth]{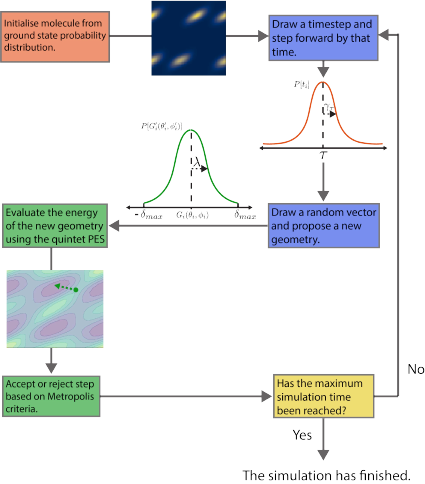}
    \caption{The basic steps of the Metropolis Monte Carlo model.}
    \label{fig:monte carlo flow}
\end{figure}

The Metropolis Monte Carlo algorithm was applied to \textbf{BP1} by linearly interpolating the two dimensional ground and quintet PES' (\cref{sec:PES}) into continuous surfaces. With these surfaces obtained, simulations were run as follows:
\begin{enumerate}
    \item Using the BP1 ground state PES, determine the ground state probability distribution for geometries at input temperature $T$ by calculating the Boltzmann distribution over a $100x100$ square grid. This provides $10000$ initial geometries to choose from with respective probabilities $p_i$
    \begin{align}
    p_i = \frac{e^{-\epsilon_i/K_BT}}{\sum_{j=1}^N e^{-\epsilon_j/K_BT}}. \label{eq:boltzmann}
\end{align}
For a system at thermal equilibrium, \cref{eq:boltzmann} describes the probability ($p_i$) the molecule is in the $i$th of $N$ possible conformations at a given time\cite{Landau2005APhysics}. Similarly, $\epsilon_i$ and $\epsilon_J$ are the energies of the $i$th and $j$th configuration respectively. $K_B$ is the Boltzmann constant and $T$ is the temperature of the system.
    \item Choose an initial geometry $\bm{X}_i(\theta_i,\phi_i)$ by drawing from the ground state probability distribution. 
    \begin{itemize}
        \item  The selection can be biased to model particular situations. For example, to model the 'freezing in' of the bulky chromophores by the solvent, the molecule initial conformations can be confined to $\theta_i-R<\theta<\theta_i + R$.
        \end{itemize}
    \item Wait an amount of time $t_i$ Lorentz distributed around the characteristic time $\tau$ with scale factor $\gamma_{\tau}$.
    \item Randomly choose a new state to step to $\bm{X}_i'(\theta_i',\phi_i') = \bm{X}_i(\theta_i +\delta\theta,\phi_i+\delta\phi)$. This is done by drawing a vector which has orientation uniformly chosen between $0\degree$ and $360\degree$, and which has magnitude drawn from a Lorentz distribution centered on zero with scale factor $\lambda$. The magnitude of the vector cannot be larger than $\delta_{max}$.
    \item Using the quintet PES, calculate the energy change $\Delta U$ which results from this displacement. If $\Delta U < 0$ the move is accepted, $\bm{X}_{i+1}(\theta_{i+1},\phi_{i+1}) = \bm{X}_i'(\theta_i',\phi_i')$. Return to step (2). If the simulation time is greater than $t_{max}$, end the simulation.
    \item If $\Delta U > 0$, choose a random number $\eta$ uniformly from $0 < \eta < 1$
    \item If $\eta < e^{-\Delta U/k_BT}$, accept the move. Here $T$ is the temperature and $k_B$ is the Boltzmann constant. Else if $\eta \geq e^{-\Delta U/k_BT}$, reject the move and $\theta_{i+1} = \theta_i$. Return to step (2). If the simulation time is greater than $t_{max}$, end the simulation.
    \item Map the resulting trajectory onto the exchange surface to determine $J$ as a function of Monte Carlo time.
\end{enumerate}
The relevant parameters for this algorithm are described in \cref{tab:model params}. A flowchart of the steps involved is depicted in \cref{fig:monte carlo flow}.
\begin{table}[ht]
\centering
\begin{tabular}{ | m{2cm} | m{13cm}| m{13cm} | } 
\hline
\textbf{Parameter}   & \textbf{Application To Model}                                       \\
\hline
$\boldsymbol{T}$   &  The temperature the simulation is run at. The temperature determines the ground state probability distribution and the Metropolis acceptance ratio $e^{-\Delta U/k_BT}$. Thus, the resulting dynamics of the model are strongly dependent on $T$.                \\
\hline
$\mathbf{t_{max}}$ & The maximum simulation time. This determines how many steps will be taken before the simulation finishes. It should be chosen to ensure that enough Monte Carlo time is simulated for the dynamics of interest (transitions between local minima) to be observed. It is important to note that this is Monte Carlo time and relating this timescale to physical timescales must be done by comparing the simulation behaviour to physical observables or other molecular dynamics models.\\
\hline
$\boldsymbol{\tau}$   & The location parameter for the Lorentz distribution of time between steps. $\tau$ represents the time taken to absorb or emit a phonon. This value should be inversely proportional to temperature\cite{Tiana2007UseChains} ensuring that a hot molecule moves frequently relative to a cold molecule. As only the ratio $\frac{\tau}{t_{max}}$ impacts the dynamics observed, $\tau$ was arbitrarily set to $\frac{100}{T}$ for all simulations. This means that at 100 K the molecule will step on average every 1 unit of time, at 200 K every $\frac{1}{2}$ units of time and so on.         \\
\hline
$\boldsymbol{\gamma_\tau}$   &  The scale factor of the Lorentz distribution with location parameter $\tau$ describing time steps.             \\
\hline
$\boldsymbol{\lambda}$   &  The scale factor for the the Lorentz distribution describing the size of steps in configuration space. The distribution is centered on zero and this factor determines the length scale over which the molecule regularly steps (see \cref{fig:monte carlo flow}). This parameter implicitly models the characteristics of the phonon bath and interactions of the molecule with the solution. If $\lambda$ is small, large steps are rare. this means the absorption or release of large phonons required for large changes in geometry are rare.                    \\
\hline
$\boldsymbol{\delta_{max}}$   & The maximum angular distance the molecule is allowed to move in one step. This determines whether large hops across the space are allowed or if the dynamics are constrained to a 'nearest neighbour' model of traversing the space where only small individual steps are allowed.                   \\
\hline
$\mathbf{R}$   & The restriction parameter. This parameter is used to model the impact of restricting the chromophore-chromophore degree of freedom ($\theta$) to a maximum amplitude of $R\degree$ from the value it was initialised in, only allowing the full rotation of the central bridge ($\phi$). This is a restriction that is expected to occur in frozen solution and will be probed to determine its impact on experimental observables. This is not applicable to the 1D model.     \\
\hline
\end{tabular}
\caption{The parameters of the Metropolis Monte Carlo model.}
\label{tab:model params}
\end{table}
\clearpage 

\subsection{Code for Metropolis Monte Carlo simulation}
\begin{lstlisting}[language=Python, caption=Metropolis Monte Carlo simulation for 2-D surfaces,label={lst:2D MC}]
#############################################################################
b1 = np.linspace(AMIN,AMAX,ANUM)
b2= np.linspace(BMIN,BMAX,BNUM)
J = interpolate.RectBivariateSpline(b1[:], b2[::2], JDATA[:,::2],kx=1,ky=1)
E = interpolate.RectBivariateSpline(b1,b2,Q_Es,kx=1,ky=1) # define interpolated energy surface to move on
G = interpolate.RectBivariateSpline(b1,b2,G_Es) # define ground state energy surface
#############################################################################
def find_pop_density(x,y,T):
    K = 1.380649e-23 #define boltzmann constant
    sump = 0
    for a in x:
        for b in y:
            sump+= (np.exp(-(G(a,b))/(K*T)))[0][0]
    G_prob = (np.exp(-(G(x,y))/(K*T))/sump)
    return G_prob

def find_pop_densityEx(x,y,T):
    K = 1.380649e-23 #define boltzmann constant
    sump = 0
    for a in x:
        for b in y:
            sump+= (np.exp(-(E(a,b))/(K*T)))[0][0]
    E_prob = (np.exp(-(E(x,y))/(K*T))/sump)
    return E_prob

def prob(E0,E1,T):
    K = 1.380649e-23
    return np.exp(-(E1-E0)/(K*T))

def find_nearest(array, value):
        array = np.asarray(array)
        idx = (np.abs(array - value)).argmin()
        return array[idx]
#############################################################################
def runsim(deltamax,vi = init ,T = temp, sim_time = tmax, tstep = twidth, runs = N, xw = width, Trajplot = True, trajnum = Nt, densplot=True, Jsurf=True,
          Jt=True,restricted=[False,None,None]):
    #Define the 2D mesh
    x, y = np.linspace(AMIN,AMAX,100), np.linspace(BMIN,BMAX,100)
    X,Y = np.meshgrid(y,x)
    theta_grid = [x,y]
    theta_gridr=[[],[]]
    theta_grid[0] = theta_grid[0].tolist()
    theta_grid[1] = theta_grid[1].tolist()
    gridsize = 180/len(theta_grid[0]) #define the increment between grid points
    xwidth = xw/gridsize   #define average increment to step in grid space by the average distance specified by xw
    dens = find_pop_density(x,y,T)
    
    t = 0 #initialise time =0
    deltat=0
    K = 1.380649e-23 #define boltzmann constant
    tau = tstep # set average wait time between steps (was 1e-12sec) now written in nanosec
    twidth = 0.5 # Distribution varince aroun which tau is centred
    tlist = []   #setup arrays to store time, angles and change in angles 
    thetas = [[],[]]
    deltas=[[],[]]
    
    a = []
    traj= [] #create a list to store trajectories in
    l = 0 #initialise tlist length to zero
    starts=[[],[]]
    for i in range(0,runs):
        if vi == 'rand':
            xi, yi = random.choice(theta_grid[0]), random.choice(theta_grid[1])
            x0, y0 = xi, yi
            starts[0].append(xi)
            starts[1].append(yi)
        elif vi == 'mins':
            ch = random.choices(dens.ravel(),weights= dens.ravel())
            coord = np.where(dens==ch)
            x0,y0 = x[coord[0]],y[coord[1]]
            starts[0].append(x0)
            starts[1].append(y0)
            
        else:
            x0, y0 = find_nearest(theta_grid[0], vi[0]), find_nearest(theta_grid[1], vi[1]) #map arbitray start points
        
        #Limit grid according to xlim and ylim in restricted argument
        if restricted[0] == True and restricted[1]!= None:
            xmin =  x0 - restricted[1]
            xmax = x0 + restricted[1]
            if xmax>180:
                xmax = xmax -180
            if xmin<0:
                xmin = xmin +180
        if restricted[0] == False or restricted[1]== None:
            xmin = 0
            xmax = 180
        if restricted[0] == True and restricted[2]!= None:    
            ymin =  y0 - restricted[2]
            ymax = y0 + restricted[2]
            if ymax>180:
                ymax = ymax -180
            if ymin<0:
                ymin = ymin +180
        if restricted[0] == False or restricted[2]== None:
            ymin = 0
            ymax = 180
            
        thetas[0].append(x0)
        thetas[1].append(y0)
        tlist.append(0)
        t = 0
        c = 0
        rjct = [[],[]]
        while t < sim_time:
            deltat=0
            while deltat <=0 or deltat>50*tau/T:
                deltat = stats.cauchy.rvs(loc=tau/T, scale=twidth, size=1)[0]  
            t += deltat  #step forward a random amount of time to interact with bath
            tlist.append(t)
            E0 = E(x0,y0)
            J0 = J(x0,y0)

            angle = random.uniform(0, 2.0*np.pi)
            mag=1000
            while abs(mag) > deltamax:
                mag = stats.cauchy.rvs(loc=0, scale=xw, size=1)[0] # 
            #mag = stats.cauchy.rvs(loc=0, scale=xw, size=1)[0]
            vector = [mag * np.cos(angle), mag * np.sin(angle)]  # scaling factor xw

            x1 = x0 +vector[0]
            y1 = y0 +vector[1]
            
            #This section updates the index of each angle, evaluates the step size and finally updates the new coordinates
            #for after the step
            f=0
            g=0
        
            while x1 > 180:
                x1 = x1 - 180
                f=1
            while x1 < 0:
                x1 = 180 + x1
                f=2
            
            if f == 1:
                delta1 = 180 - abs(x0-x1)
            elif f== 2:
                delta1 = -(180 - abs(x0-x1))
            elif f == 0:
                delta1 = x0 - x1
                
            while y1 > 180:
                y1 = y1 - 180
                g=1
            while y1 < 0:
                y1 = 180 + y1
                g=2
        
            if g == 1:
                delta2 = 180 - abs(y0-y1)
            elif g== 2:
                delta2 = -(180 - abs(y0-y1))
            elif g == 0:
                delta2 = y0 - y1
                
            #Here we evaluate the energy of the step and implement the metropolis algorithm  
            E1 = E(x1,y1)
            if xmin<xmax and xmin>x1 or x1>xmax :
                E1 = 1000000000
            if xmin>xmax and x1>xmax and x1<xmin:
                E1 = 1000000000
            if ymin<ymax and ymin>y1 or y1>ymax :
                E1 = 1000000000
            if ymin>ymax and y1>ymax and y1<ymin:
                E1 = 1000000000
            pcut = random.random()
            if prob(E0,E1,T) > pcut:
                x0 = x1
                y0 = y1
                a.append(1)
            elif prob(E0,E1,T) <= pcut:
                delta = 0
                a.append(0)
                rjct[0].append(x1)
                rjct[1].append(y1)
            thetas[0].append(x0)
            thetas[1].append(y0)
            deltas[0].append(delta1)
            deltas[1].append(delta2)
            
        if i < trajnum:
            traj.append([tlist[l:],thetas[0][l:], thetas[1][l:], rjct[0][l:],rjct[1][l:]])
            l = len(tlist)
#############################################################################
\end{lstlisting}
\clearpage
\section{Spin Dynamics Calculations}
\label{sec:SI-Spin calcs}
 To investigate the ensemble spin dynamics that evolve from Monte Carlo \textbf{BP1} trajectories, we explored the impact of changes in temperature, hopping rate and chromophore restriction. The simulation length $t_{max}$ was fixed at $20 000$ units of Monte Carlo time. This was chosen to allow a number of transitions to occur between local energy minima within each simulation. The full procedure is listed below and the relevant code for the simulation of these dynamics is found in Code Excerpt \ref{lst:QuTip model}.
\begin{enumerate}
    \item Generate 50 molecular dynamics trajectories using the Metropolis Monte Carlo model at temperature $T$ and with chromophore restriction $R$. The remaining parameters were fixed according to \cref{tab:largesim params QT}.
    \item For each trajectory, attain $J(t)$ and $\theta(t)$ by linearly interpolating the values of $J$ and $\theta$ between each time step and scaling the units of time such that $t_{max}=$ Timescale.
    \item Solve the time-dependent Schr\"odinger equation with spin Hamiltonian,
    \begin{align}
        &\hat{H}(t) = \hat{H}_Z + \hat{H}_{ZFS}(t) +\hat{H}_{EX}(t),\\
        &\hat{H}_Z = \mu_B g \sum_{i=1,2}\sum_{j=x,y,z} B_j\hat{S}_{ij},\\
        &\hat{H}_{ZFS}(t) = \sum_{i=1,2} \hat{S}_i \cdot \mathbf{D_i}\left(\theta(t)\right) \cdot \hat{S}_i, \\
        &\hat{H}_{EX}(t) = J(t)\hat{S}_1\cdot\hat{S}_2,
    \end{align} using the \verb+mesolve+ function of QuTip. The parameters that define this spin Hamiltonian are listed in\cref{tab:largesim params QT}. This step is the most computationally expensive. To make the many simulations feasible, the time-dependent functions are only sampled by QuTip at 100 points as the state evolves.
    \item Repeat steps (2) and (3) for each timescale in \cref{tab:largesim params QT}.
    \item Repeat steps (1) through (4) for all combinations of the temperature (T) and restriction (R) parameters in \cref{tab:largesim params}.
\end{enumerate}
\newpage
\begin{table}[ht]
\centering
\begin{tabular}{ | m{2cm} | m{5cm}| m{5cm} | } 
\hline
\textbf{Parameter}   & \textbf{Value}                                       \\
\hline
${T}$   &  10, 50, 100, 200, 300, 500 ($\degree K$)               \\
\hline
${t_{max}}$ & 20000\\
\hline
${\tau}$   & $\frac{100}{T}$\\
\hline
${\gamma_\tau}$   &    0.5           \\
\hline
${\lambda}$   & $1\degree$                    \\
\hline
${\delta_{max}}$   & $180\degree$               \\
\hline
${R}$   & None, $10\degree$    \\
\hline
\end{tabular}
\label{tab:largesim params}
\begin{tabular}{ | m{3cm} | m{3cm}| m{3cm} | } 
\hline
\textbf{Parameter}   & \textbf{Value}                                       \\
\hline
${g_x=g_y=g_z}$   &  2.002               \\
\hline
${D}$ & 1138 MHz\\
\hline
${E}$   & 19 MHz\\
\hline
$\boldsymbol{B}=(B_x,B_y,B_z)$   &   (0,0,350mT)           \\
\hline
{Timescale} & 10, 100, 1000 (ns)\\
\hline
\end{tabular}\label{tab:largesim params QT}
\caption{The parameters of the Monte Carlo (left) and spin dynamics (right) simulations.}
\end{table}
\begin{table}[ht]
\centering
\begin{tabular}{ | m{4cm} | m{8cm}| m{8cm} | } 
\hline
\textbf{Resonance Frequency of $J$ (GHz)} & \textbf{Avoided Crossing(s)}  \\
\hline
$0.00$ & $^1(TT)_{0} \sim {^3(TT)_{0}} \sim {^5(TT)_{0}}$\newline$^3(TT)_{\pm1} \sim {^5(TT)_{\pm1}}$ \\
\hline
$3.27$ & $^1(TT)_{0}\sim{^5(TT)_{-1}}$ \\
\hline
$4.90$ & $^3(TT)_{0}\sim{^5(TT)_{-1}}$ \newline $^3(TT)_{+1}\sim{^5(TT)_{0}}$ \newline $^3(TT)_{-1}\sim{^5(TT)_{-2}}$ \\
\hline
$6.54$ & $^1(TT)_{0}\sim^5(TT)_{-2}$ \\
\hline
$9.81$ & $^1(TT)_{0}\sim{^3(TT)_{-1}}$\newline $^3(TT)_{0}\sim{^5(TT)_{-2}}$ \newline $^3(TT)_{+1}\sim{^5(TT)_{-1}}$ \\
\hline
$14.71$ & $^3(TT)_{1}\sim{^5(TT)_{-2}}$ \\
\hline
\end{tabular}
\caption{The magnitude of $J$ at which the avoided crossings between spin states, denoted by $\sim$, occur.}
\label{tab:resonances}
\end{table}

\begin{lstlisting}[language=Python, caption={Relevant code for the construction of the Qutip Model}, label={lst:QuTip model}]
#############################################################################
#set up the qutip hamiltonians / quantum objects
#set constants to units of Gz (hence time is in ns)
g = 2.002
D_zfs = 1138e-3 #in units of GHz
E_zfs = 19e-3  #in units of GHz
B = 350*10**(-3) #in units of Tesla
mu = scp.physical_constants['Bohr magneton in Hz/T'][0]*10**(-9) #GHz/T
#############################################################################
#Define I of spin-1 hilbert space
I = qt.qeye(3)
#define pauli matrices for spin-1 hilbert space
sx,sy,sz = qt.jmat(1,'x'),qt.jmat(1,'y'),qt.jmat(1,'z')
#define spin operators for triplet pair
sx2 = qt.tensor(sx,I) + qt.tensor(I,sx) # spin projection operators
sy2 = qt.tensor(sy,I) + qt.tensor(I,sy)
sz2 = qt.tensor(sz,I) + qt.tensor(I,sz)
S2 = sx2**2 + sy2**2 + sz2**2  # Total spin S^2
#############################################################################
# define our 9 eigenstates of S^2 to project onto
states = (S2 + sz2).eigenstates()[1]
psi0 = states[0]
S0 = states[0]*qt.dag(states[0])
Tmin1 = states[1]*qt.dag(states[1])
T0 = states[2]*qt.dag(states[2])
Tplus1 = states[3]*qt.dag(states[3])
Qmin2 = states[4]*qt.dag(states[4])
Qmin1 = states[5]*qt.dag(states[5])
Q0 = states[6]*qt.dag(states[6])
Qplus1 = states[7]*qt.dag(states[7])
Qplus2 = states[8]*qt.dag(states[8])
#############################################################################
#Define exchange hamiltonian (sans J(t))
Hex = (qt.tensor(sz,sz) + qt.tensor(sx,sx) + qt.tensor(sy,sy))
#define ZFS tensor
zfst = [[E_zfs-D_zfs/3,0,0],[0,-E_zfs-D_zfs/3,0],[0,0,2*D_zfs/3]] #for molecule along z axis (eul = 0)
#############################################################################
#Define the function that takes a MC trajectory and timescale as input and outputs spin densities over #time
def qutip_evol(trajectory,qutip_time,new_name):
    #start = time.perf_counter()
    #Set time grid
    t= np.linspace(0 ,qutip_time-0.01,100)
    S0_t = []
    Q0_t = [] 
    #Interpolate J(t) and theta(t)
    theta_t = interpolate.interp1d(trajectory[0], np.array(trajectory[2]).squeeze()*np.pi/180,kind='linear')
    Js = J.ev(trajectory[1],trajectory[2]).squeeze()
    
    #smoothing option to reduce cost of noisy signals
    #N=10000
    #Jsmooth=uniform_filter1d(Js, size=N)
    #evaluate = interpolate.interp1d(np.array(trajectory[0]), Jsmooth.squeeze(),kind='linear')
    evaluate = interpolate.interp1d(np.array(trajectory[0]), J.ev(trajectory[1],trajectory[2]).squeeze(),kind='linear')
    #define our function J(t) as the interpolated function  
    def J_t(t,args):
        return evaluate(t)
    #Define rotation matrix for relative angle of two chromophores
    def R_rel(z1,y,z2):
        Rz1 = np.matrix([[np.cos(z1),-np.sin(z1),0],[np.sin(z1),np.cos(z1),0],[0,0,1]])
        Ry = np.matrix([[np.cos(y),0,np.sin(y)],[0,1,0],[-np.sin(y),0,np.cos(y)]])
        Rz2 = np.matrix([[np.cos(z2),-np.sin(z2),0],[np.sin(z2),np.cos(z2),0],[0,0,1]])
        r = np.matmul(Rz2,Ry,Rz1)
        return r

    #Define rotation matrix for molecule in relation to B-field
    def R_B(z1,y,z2):
        Rz1 = np.matrix([[np.cos(z1),-np.sin(z1),0],[np.sin(z1),np.cos(z1),0],[0,0,1]])
        Ry = np.matrix([[np.cos(y),0,np.sin(y)],[0,1,0],[-np.sin(y),0,np.cos(y)]])
        Rz2 = np.matrix([[np.cos(z2),-np.sin(z2),0],[np.sin(z2),np.cos(z2),0],[0,0,1]])
        r = np.matmul(Rz2,Ry,Rz1)
        return r

    #define total rotation matrix as combo of both
    def erot(eul):
        return np.matmul(R_rel(eul[0],eul[1],eul[2]),R_B(eul[3],eul[4],eul[5]))

    #define time dependent elements of the rotation matrix
    def R00(t,args):
        r = np.matmul(np.transpose(erot([0,theta_t(t),0,0,0,0])),np.matmul(zfst,erot([0,theta_t(t),0,0,0,0])))
        return r[0,0]
    def R11(t,args):
        r = np.matmul(np.transpose(erot([0,theta_t(t),0,0,0,0])),np.matmul(zfst,erot([0,theta_t(t),0,0,0,0])))
        return r[1,1]
    def R22(t,args):
        r = np.matmul(np.transpose(erot([0,theta_t(t),0,0,0,0])),np.matmul(zfst,erot([0,theta_t(t),0,0,0,0])))
        return r[2,2]
    def R01(t,args):
        r = np.matmul(np.transpose(erot([0,theta_t(t),0,0,0,0])),np.matmul(zfst,erot([0,theta_t(t),0,0,0,0])))
        return r[0,1]
    def R02(t,args):
        r = np.matmul(np.transpose(erot([0,theta_t(t),0,0,0,0])),np.matmul(zfst,erot([0,theta_t(t),0,0,0,0])))
        return r[0,2]
    def R12(t,args):
        r = np.matmul(np.transpose(erot([0,theta_t(t),0,0,0,0])),np.matmul(zfst,erot([0,theta_t(t),0,0,0,0])))
        return r[1,2]

    # Construct ZFS hamiltonian with time dependent elements
    def zfs(zfst,eul):
        sx,sy,sz = qt.jmat(1,'x'),qt.jmat(1,'y'),qt.jmat(1,'z')
        R = np.matmul(np.transpose(erot(eul)),np.matmul(zfst,erot(eul)))
        H = R[0,0]*(sx**2)+R[1,1]*(sy**2)+R[2,2]*(sz**2)+R[0,1]*(sx*sy+sy*sx)+R[0,2]*(sx*sz+sz*sx)+R[1,2]*(sy*sz+sz*sy)
        return H

    #Here we have created a time evolving quantum object by making the rotation matrix time dependent
    #as it stands this only changes the relative orientation of one chromophore and keeps B along z- axis
    Hzfs1 = qt.QobjEvo([[qt.tensor(sx**2,I), R00], [qt.tensor(sy**2,I),R11], [qt.tensor(sz**2,I),R22], 
                        [qt.tensor(sx*sy+sy*sx,I),R01], [qt.tensor(sx*sz+sz*sx,I),R02], [qt.tensor(sy*sz+sz*sy,I),R12]])
    
    Hzfs = Hzfs1 + qt.tensor(I,zfs(zfst,np.zeros(6))) # we pick one zfs tensor to be our axis and rotate                                                  # the other by the relative orientation of the two.
    
    #define Zeeman hamiltonian
    Hz = mu*B*g*(qt.tensor(sz,I) + qt.tensor(I,sz))
    Hz = mu*B*g*(qt.tensor(sz,I) + qt.tensor(I,sz))
    #Define exchange hamiltonian
    Hex1 = qt.QobjEvo([[Hex,J_t]])
    
    #Define total hamiltonian as sum of zeeman, zfs and exhange
    Ht = Hz + Hzfs + Hex1
    
    #define time evolved state as a func of time
    options= qt.Options(nsteps=8000)
    state_t=qt.mesolve(Ht,psi0,t,options=options).states
    return t,state_t,theta_t,J_t
\end{lstlisting}
\end{document}